\documentclass[sn-mathphys-num]{sn-jnl}

\usepackage{graphicx}%
\usepackage{multirow}%
\usepackage{amsmath,amssymb,amsfonts, physics}%
\usepackage{amsthm}%
\usepackage{mathrsfs}%
\usepackage[title]{appendix}%
\usepackage{xcolor}%
\usepackage{textcomp}%
\usepackage{manyfoot}%
\usepackage{booktabs}%
\usepackage{algorithm}%
\usepackage{algorithmicx}%
\usepackage{algpseudocode}%
\usepackage{listings}
\usepackage{multirow}%

\theoremstyle{thmstyleone}%
\theoremstyle{thmstyletwo}%

\theoremstyle{thmstylethree}%

\usepackage{diagbox}
\usepackage{tikz}
\newcommand{\circled}[1]{\tikz[baseline=(char.base)]{\node[shape=circle,draw,inner sep=1pt] (char) {#1};}}

\usetikzlibrary{calc,positioning, arrows}

\newcommand{\lrfb}[1]{\left(#1\right)}

\begin{document}
	
	\title[Article Title]{Teleportation Fidelity of  Binary Tree Quantum Repeater Networks}
	
	
	\author*[1]{\fnm{Soumit} \sur{Roy}\orcid{0009-0002-4579-0491}}\email{soumit.roy@research.iiit.ac.in}
	
	\author[2]{\fnm{Md Rahil} \sur{Miraj}\orcid{0009-0002-8299-3589}}\email{rahilmiraj@gmail.com}
	
	\author[3]{\fnm{Chittaranjan} \sur{Hens}\orcid{0000-0003-1971-089X}}\email{chittaranjan.hens@iiit.ac.in}
	
	\author[1]{\fnm{Ganesh} \sur{Mylavarapu}\orcid{0000-0002-9242-9896}}\email{ganesh.mylavarapu@research.iiit.ac.in}
	
	\author[4]{\fnm{Subrata} \sur{Ghosh}\orcid{0000-0003-4043-1422}}\email{jaba.subrata94@gmail.com}
	
	\author[1,5]{\fnm{Indranil} \sur{Chakrabarty}\orcid{0009-0001-0415-0431}}\email{indranil.chakrabarty@iiit.ac.in} 
	
	\affil*[1]{\orgdiv{Centre for Quantum Science and Technology}, \orgname{International Institute of Information Technology, Hyderabad}, \orgaddress{\city{Hyderabad}, \postcode{500032}, \state{Telangana}, \country{India}}}
	
	\affil[2]{\orgdiv{Theoretical Statistics and Mathematics Unit}, \orgname{Indian Statistical Institute, Kolkata}, \orgaddress{\city{Kolkata}, \postcode{700108}, \state{West Bengal}, \country{India}}}
	
	\affil[3]{\orgdiv{Center for Computational Natural Science and Bioinformatics}, \orgname{International Institute of Information Technology, Hyderabad}, \orgaddress{\city{Hyderabad}, \postcode{500032}, \state{Telangana}, \country{India}}}
	
	\affil[4]{\orgdiv{Division of Dynamics}, \orgname{Lodz University of Technology}, \orgaddress{\city{Stefanowskiego 1/15, 90-924}, \state{Lodz}, \country{Poland}}}
	
	\affil[5]{\orgdiv{Center for Security, Theory and Algorithmic Research}, \orgname{International Institute of Information Technology, Hyderabad}, \orgaddress{\city{Hyderabad}, \postcode{500032}, \state{Telangana}, \country{India}}}
	
	
	\abstract{Binary tree network, being a subclass of Cayley tree network, is a significant topological structure used for information transfer in a hierarchical sense. In this article, we consider four types of binary tree repeater networks (directed and undirected, asymmetric and symmetric) and obtain the analytical expressions of the average of the maximum teleportation fidelities for each of these binary tree networks. We contribute a methodology for the analytical calculation of pathlengths in all considered graph types. Based on these, we have used simple Werner state-based models and are able to identify the parameter ranges for which these networks can show quantum advantage. We also explore the role of maximally entangled states in the network to enhance the quantum advantage. We provide a detailed examination of the large-scale behavior of these networks, obtaining the limiting value of the average maximum teleportation fidelity as the number of nodes, $N$, approaches infinity, same as fractal tree. Our findings reveal that the directed symmetric binary tree represents the most advantageous topology for quantum teleportation within this context. From the context of quantum repeater networks, this work makes a significant advancement in the process of identifying resourceful tree networks for distributed quantum teleportation \textit{i.e.} teleportation between all possible sources and targets.}


	\maketitle
	
	\section{Introduction}
	In the last decade, we have seen a steady rise in the research and implementation of quantum networks with the vision of creating a full-stack quantum internet \cite{dowling2003quantum, kimble2008quantum, Google, Rigeti, long2022, Yan2021}. Quantum networking also becomes crucial in distributed quantum computing, where separated quantum computers can work on a similar algorithm \cite{aaronson2011computational,menicucci2006universal,knill2001scheme,cirac1999distributed,fitzi2001quantum,giovannetti2008quantum,das2021practically, bollinger1996optimal,giovannetti2006quantum,pironio2010random,yang2019distributed,sazim2013study}. Not only computing, but quantum networks can become a test bed for various information processing tasks with many more parties and distributed scenarios. In that spirit, we have seen network topology to play a significant role and value-addition in choosing a network best suited for a particular task \cite{mylavarapu2024teleportation}. Unlike a classical network \cite{albert2002statistical,watts1998collective,pastor2015epidemic,gao2016universal,boccaletti2014structure,ji2023signal,hens2019spatiotemporal,ghavasieh2024diversity,meena2023emergent,moore2020predicting}, one of the unexplored areas is the large-scale quantum networks. For a practical purpose, quantum networks can be deployed either through satellite-based free space technology \cite{aspelmeyer2003long,bedingtonprogress} or large ground-based repeater technology \cite{briegel1998quantum,duan2001long,gour2004remote,sazim2013study}. In an entanglement-based repeater network, entanglement, being the key ingredient, have a crucial role in various communication tasks like sending quantum information, classical information, key generation \cite{BENNETT20147, ekert1991quantum, B92} and secret sharing \cite{bennett1993teleporting,horodecki1996teleportation,bennett1992communication,srivastava2019one,vempati2021witnessing,patro2017non,pati2000minimum,bennett2005remote,ekert1991quantum,shor2000simple,sohail2023teleportation,adhikari2010probabilistic,singh2024controlled,hillery1999quantum,ray2016sequential,sazim2015retrieving}. In such networks the distribution of entanglement (both bipartite \cite{gour2004remote,sazim2013study} and multipartite \cite{meignant2019distributing})  becomes a crucial factor despite limitations of achieving relevant quantum advantages \cite{neumann2025no,bauml2015limitations,das2021practically}.
	In this article, we consider a ground-based repeater network specially curated for the purpose of teleportation.
	
	\par Quantum teleportation \cite{bennett1993teleporting}, being one of the fascinating discoveries of the last century, enables us to send quantum information from one location to another. However, the transfer of information is not limited to unknown qubits and can also be extended to the coherence of unknown quantum states \cite{sohail2023teleportation}. Though this can be achieved with entangled states, not all entangled states are useful for teleportation \cite{horodecki1996teleportation,chakrabarty2010teleportation,chakrabarty2011deletion,adhikari2008quantum} and hence the existence of a separate teleportation witness \cite{ganguly2011entanglement} to identify states that are useful for teleportation. It is interesting to note that there are states that cannot be useful for teleportation even after the application of the global unitary operators \cite{patro2022absolute}.  The ability of a two-qubit state to be useful for a teleportation resource is given by the teleportation fidelity $F^{\text {tel}}$, which is a function of the correlation matrix $T$ of the two-qubit state \cite{horodecki1996teleportation}. If $F^{\text{tel}}>2/3$ (the classical limit without any entangled state), then there is a quantum advantage present in the state for carrying out the process of teleportation. For a simple Werner state (parametrized by $p$) as the resource state, the teleportation fidelity becomes a function of $p$ and becomes useful for teleportation only when $p>1/3$. 
	
	\par
\begin{figure*}[htbp]
		\centering
		
		\begin{minipage}[t]{0.45\textwidth}
			\centering
			\includegraphics[width=\textwidth]{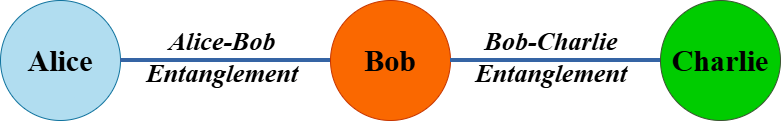}
			\vspace{1mm}
			(a)
			\label{fig:es_a}
		\end{minipage}
		\hfill
		%
		\begin{minipage}[t]{0.45\textwidth}
			\centering
			\includegraphics[width=\textwidth]{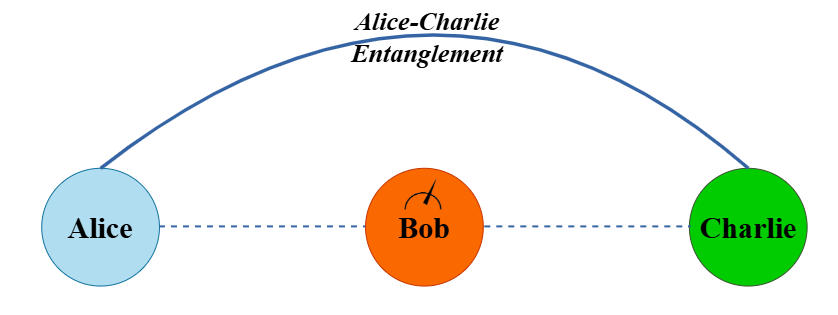}
			\vspace{1mm}
			(b)
			\label{fig:es_b}
		\end{minipage}
		\hfill \\
		
		\hfill
		\begin{minipage}[t]{0.5\textwidth}
			\centering
			\includegraphics[width=\textwidth]{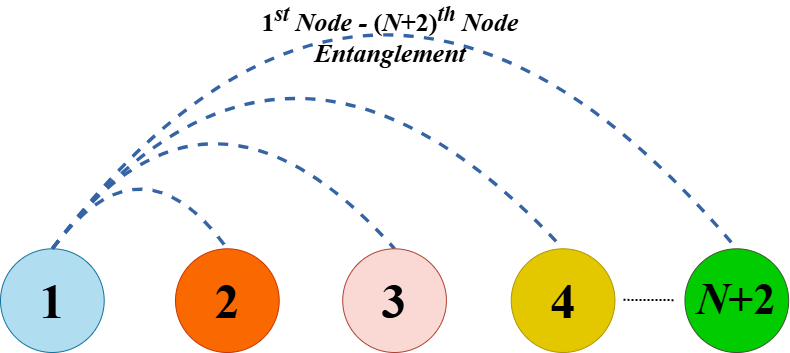}
			\vspace{1mm}
			(c)
			\label{fig:es_c}
		\end{minipage}
		\hfill
		\caption{
			Entanglement Swapping: 
			(a) Alice and Charlie are not connected, but entangled pairs are shared between Alice-Bob and Bob-Charlie. 
			(b) Bob's measurement creates entanglement between Alice and Charlie. 
			(c) Sequential protocol to establish entanglement between $N+2$ nodes using $N$ swaps.
		}
		\label{fig:ES}
	\end{figure*}

 In a repeater-based quantum network scenario, establishment of entanglement between two distant nodes \cite{gour2004remote,guerra2025entanglement,sazim2013study} happens through the process of entanglement swapping in the repeater stations \cite{mylavarapu2023entanglement}. Given the fidelity of intermediate states, it becomes interesting to find out the teleportation fidelity of the entangled state established between the source and target as a result of swapping at the repeater station (Fig. \ref{fig:ES}). In a quantum repeater network made up of Werner states (parametrized by a set of values $p_i$), the maximum teleportation fidelity through a particular path (P) connecting the source (S) and
	the target (T) can be calculated $F^{\text{tel}}$ as $F^{\rm max}_{{\rm ST},\mathcal P}(\rho_{\rm wer}) = \left(1 + \prod_{i\in\mathcal P} p_{i}\right)/2$ \cite{mylavarapu2023entanglement,sen2005entanglement}. In fact, this is the teleportation fidelity of the final entangled state between S and T created by swapping through that path. In general, in a full-scale quantum network, there can be multiple paths from a source to a target, and hence multiple teleportation fidelities between S and T based on different paths can exist. However, we take the maximum of all the teleportation fidelities when we fix the source and target \cite{mylavarapu2024teleportation}. To quantify the resourcefulness of a network globally in the context of teleportation, a measure called \textit{average of maximum teleportation fidelity} ($F^{\text{tel}}_{\text{avg}}$) achieved by all paths was introduced in a recent work \cite{mylavarapu2024teleportation}. 
	This has given us a global metric ($F^{\text{tel}}_{\text{avg}}$) to estimate the capability of a network in carrying out teleportation over any source and target, and eventually ranking them.  In that article \cite{mylavarapu2024teleportation}, the authors have investigated how teleportation fidelity (on an average scale) varies in both loop and loopless networks. We have examined tree sequences that range from chains with all nodes having degree two (apart from the extreme two nodes) to stars, where the hub has the most links. Other trees, like branching trees, that can be created by gradually rearranging nodes from the chain, have also been investigated \cite{mylavarapu2024teleportation}. 
	It's been demonstrated that the network can achieve the quantum average fidelity with a lower average distance. In the case of a repeater-based distributed quantum protocol, the intermediate state can get noisier, and in that case, the efficient protocol proposed in \cite{ghosal2025} can be used to ensure the optimal teleportation fidelity.  
	\par Motivated by this newly introduced measure $F_{\text{avg}}^{\text{tel}}$ in undirected simple trees, we focus here on a class of directed and undirected binary trees, a subclass of the Cayley tree. A Cayley tree is an acyclic, connected graph in which each node has up to $k$ children.
	The binary trees are a specific subcategory of Cayley tree. A left child and a right child are the maximum number of offspring that each node in the binary tree, a hierarchical tree structure, can have. Note that binary trees are used in computer science for designing algorithms and data structures \cite{jeang2005, ayman2011}.
	A lot of studies have been conducted to investigate the role of (un)directed trees on percolation \cite{bethe1935, Stauffer2003, Christensen2005, Nogawa2009, Minninghen2010, Bonomo2024}, 
	synchronization~\cite{Dekker2013, sarika2014},
	ecological stability, and in the immune network model \cite{LuizHenriqueDore2019, Anderson1993}.
	Investigating the average of maximum teleportation fidelity in directed and undirected binary trees is necessary to understand the process of quantum information transfer in a hierarchical network structure and, hence, will be the focus of the current work. Specifically, both symmetric and asymmetric binary trees will be used to study the $F_{\text{avg}}^{\text{tel}}$ in this article. In the symmetric situation, each node has two children (left and right) at each hop/level. The right child will not be taken into consideration if each level has an asymmetric instance. {Note that increasing the depth of an undirected binary tree can lead to a fractal tree \cite{Mandelbrot1999}. This scenario can be mapped to the undirected symmetric binary tree.} The current work unfolds two aspects: (i)  How does symmetry in a binary graph shape the global teleportation fidelity, and  (ii)  how does directionality in links influence the global behavior?
	\par In this article, we consider four types of binary trees: directed asymmetric binary tree (\textbf{DABT}), directed symmetric binary tree (\textbf{DSBT}), undirected asymmetric binary tree (\textbf{UABT}) and undirected symmetric binary tree (\textbf{USBT}), in which the USBT will be a fractal when $d$ is very large. For each case, we analytically calculate the number of shortest paths of different pathlengths. Further, based on these numbers, we analytically and numerically find out the average of maximum teleportation fidelities for all these binary tree topologies (See Section \ref{sec2d}, Fig.\ \ref{fig:bt}). Our network models are based on edges made up of Werner states parametrized by the parameter $p$ uniformly over the network. We also obtain the range of $p$ for which each networks show quantum advantage with varying network size $N(d)$, (parametrized by depth $d$). In that process, we also find out the limiting value of the average of the maximum teleportation fidelity for each of these networks for a large number of $N(d)$. Given that each Werner state is useful for teleportation for $p>1/3$, we find the lower bound of the average of the maximum teleportation fidelity for each network and capture its variation with increasing $N(d)$. 
	\par In Section \ref{sec2}, we give a brief introduction to existing concepts like entanglement swapping, average of maximum teleportation fidelity and binary tree network. In Section \ref{sec3}, we obtain the average of maximum teleportation fidelity for each of these four binary networks. In Section \ref{sec4}, we analyze the quantum advantage for each of these networks and also find the lower bounds of the average of maximum teleportation fidelities when each link is useful for teleportation. In Section \ref{sec5}, we have calculated the average of maximum fidelity when the input state parameter of the Werner states is different for each edge of the corresponding binary tree network (the input parameters are chosen from a uniform distribution). Finally, in Section \ref{sec6}, we have concluded by summarizing the findings of this work.  
	
	\section{Preliminaries} \label{sec2}
	In this section, we discuss a few preliminary but related concepts like entanglement swapping, the average of maximum teleportation fidelity and the binary tree.  
	\subsection{Entanglement Swapping and Quantum  Repeater}
	Entanglement swapping is the process by which one can teleport a quantum entanglement from one location to another, thus by connecting two nodes that are not initially entangled with each other. The simplest scheme of entanglement swapping involves three parties, Alice, Bob and Charlie. Here Alice and Charlie shares Bell state $|\Phi^+\rangle_{AB_1}$ and $|\Phi^+\rangle_{B_2C}$ with Bob. Now, if Bob carries out two two-qubit Bell-state measurements on his qubits $B_1$ and $B_2$, Alice and Charlie will end up with two two-qubit states depending upon the measurement outcomes (See Fig.\ \ref{fig:ES}) at Bob's location,
	\begin{eqnarray}
		&&|\Phi^+\rangle_{B_1B_2}\rightarrow |\Phi^+\rangle_{AC},{}\nonumber\\&&
		|\Psi^+\rangle_{B_1B_2}\rightarrow X_{C}|\Phi^+\rangle_{AC},{}\nonumber\\&&
		|\Phi^-\rangle_{B_1B_2}\rightarrow Z_{C}|\Phi^+\rangle_{AC},{}\nonumber\\&&
		|\Psi^-\rangle_{B_1B_2}\rightarrow Z_{C}X_{C}|\Phi^+\rangle_{AC}.
	\end{eqnarray}
	Here, $X$ and $Z$ are the Pauli operators. Alice and Charlie are able to generate a maximally entangled state between them without interacting directly. This process can be explored to create a long-distance entanglement between two parties which are not initially connected. Now, if instead of a maximally entangled state we have the same  Werner state $\rho_{\text wer}$ with input state parameter $p$,
	\begin{equation}
		\rho_{\text{wer}}=\frac{1-p}{4}\mathbb{I}+p\dyad{\Psi^{-}},
		\label{eq:wener}
	\end{equation}
	as a resource between Alice-Bob and Bob-Charlie, respectively. After first swapping the state between Alice and Charlie, will become \cite{mylavarapu2023entanglement, sen2005entanglement},
	\begin{equation}
		\rho_{\text{wer}}^{(1)}= \frac{1-p^{2}}{4}\mathbb{I}+p^{2}\dyad{\Psi^{-}}.    
	\end{equation}
	If we extend this idea to a $(N+2)$ node scenario, to establish the connection between the $1$\textsuperscript{st} and the $(N+2)$\textsuperscript{th} node, then there should be $N$ such swappings (See Fig.\ \ref{fig:ES}(c)). The resultant state  will become,
	\begin{equation}
		\rho_{\text{wer}}^{(N)}= \frac{1-p^{N+1}}{4}\mathbb{I} + p^{N+1}\dyad{\Psi^{-}}.    
	\end{equation}
	\\
	The idea of entanglement swapping can be extended to implement a quantum repeater between two nodes of a quantum network.

	\subsection{ Teleportation Fidelity and Average of Maximum Teleportation Fidelity}
	
	Quantum teleportation is one of the celebrated methods by which one can send quantum information (qubit/qudit) (also quantum coherence) from one location to another with the help of an entangled state shared between the parties. If we consider a perfect teleportation protocol, we can send an arbitrary unknown qubit with a shared Bell state \cite{bennett1993teleporting}. If we do not have a maximally entangled state, we can still send a qubit with a non-maximally entangled resource \cite{agrawal2002probabilistic}. This can be achieved with a certain probability of success if not perfectly \cite{agrawal2002probabilistic}. Interestingly, it can be seen that  for a two-qubit system or state $\rho$, the capacity of the state to act as a resource for teleportation can be quantified in the form of teleportation fidelity \cite{horodecki1996teleportation}, 
	\begin{eqnarray}
		\mathcal{F}=(1+\mathcal{N}(\rho)/3)/2,
	\end{eqnarray}
	with $\mathcal{N}(\rho)=\Tr(\sqrt{T^{\dagger}T})$, where $T$ is the correlation matrix of $\rho$. The elements of the matrix $T$ are defined as  $t_{mn} = \Tr{\rho(\sigma_m\otimes \sigma_n)}$. Here $\sigma_i$ are the Pauli matrices and $(m,n)=1,2,3$. It is interesting to note that there are entangled states that are not useful for teleportation \cite{horodecki1996teleportation,chakrabarty2010teleportation,adhikari2008quantum,chakrabarty2011deletion} and hence the teleportation witness operator \cite{ganguly2011entanglement}. The idea of quantum advantage comes from the fact of how much more one can teleport than one can do classically without an entangled state. For such teleportation, the maximum fidelity comes out to be $2/3$ \cite{horodecki1996teleportation}. For a Werner state $ \rho_{\text{wer}}$, this fidelity comes out to be $\mathcal{F}=\frac{1}{2}(1+p)$.  It shows quantum advantage when $\mathcal{F}>2/3$ \textit{i.e.} when $p>1/3$.
	\par For a repeater based Werner network, the achievable teleportation fidelity through a particular path ($\mathcal P$) connecting the source (S) and
	the target (T) can be obtained as $F^{\rm tel}_{{\rm ST},\mathcal P}(\rho_{\text{wer}}) = \left(1 + \prod_{i\in\mathcal P} p_{i}\right)/2$ \cite{mylavarapu2023entanglement,mylavarapu2024teleportation}. In a full-scale quantum network, there may generally be many paths from a source to a target. But when we fix the source and target, we take the maximum of all the teleportation fidelities. However when all $p_i$ are same and is  equal to $p$ then $F^{\rm tel}_{{\rm ST},\mathcal P}(\rho_{\text{wer}}) = \frac{1+p^{N+1}}{2}$ for $N+2$ nodes scenario. \\
	%
	\indent The average of maximum teleportation fidelity for all paths in a network is a recent concept designed to reflect the ability of a network to act as a resource for carrying out teleportation between any source and any target \cite{mylavarapu2024teleportation}. 
	If S and T are connected via multiple paths, let $\mathcal P_{\rm max}$ be the path with the maximum fidelity. We get the average of highest-achievable teleportation fidelity if we take the average of $F^{\rm max}_{{\rm ST},\mathcal P_{\rm max}}(\rho)$ over all possible combinations of S and T (\textit{i.e.}, any pair of nodes can be the source and the target):
	\begin{eqnarray}
		F^{\rm max}_{\rm avg}(\rho) = \langle F^{\rm max}_{{\rm ST},\mathcal P_{\rm max}}(\rho)\rangle_{{\rm ST}}= \langle F^{\rm max}_{\mathcal P}(\rho)\rangle_{\mathcal P}, \label{eq:fmaxavgdef}
	\end{eqnarray}
	where the second step follows from the fact that in the absence of loops, the path between any S and T pair is unique. Hence, in our case, where there is no loop, averaging over S and T pairs is equivalent to averaging over all possible paths in the network. From now on we denote this average as $\langle F^{\rm max}_{\mathcal P}(\rho)\rangle_{\mathcal P}$ as $F^{\text{tel}}_{\text{avg}}$ and call it as average teleportation fidelity as there is unique fidelity between source and target in absence of loops. 
	
	
	
	\subsection{Binary Tree Network} \label{sec2d}
	A binary tree is one of the hierarchical tree data structures, with each node having at most two child nodes.Scalable and efficient network communication is achieved through the widespread use of hierarchical tree structures in communication networks for routing, multicast distribution, and efficient packet forwarding, all of which are well studied in classical network scenarios \cite{jeang2005, ayman2011}. In addition to these, the tree cluster scenarios are used in quantum repeater architecture to enable long-range quantum communication \cite{borregaard2020}, and tree-encoded photonic cluster states are utilised to enable loss-tolerant quantum memory/information storage \cite{varnava2006}. Furthermore, the binary tree serves as the basic unit of fractal tree structures, as discussed in the literature, and represents a special case of the Cayley tree. Based on the structure, there are two types of binary trees: \textit{asymmetric} and \textit{symmetric}. Similarly, based on the directionality of the edges, we can classify binary trees into two types: \textit{directed} and \textit{undirected}. Here in this article, we consider four different types of binary trees: directed asymmetric and symmetric binary trees (DABT, DSBT) and undirected asymmetric and symmetric binary trees (UABT, USBT). All other binary tree configurations can be regarded as intermediate cases lying between these four extremes. These limiting cases also illustrate how the average teleportation fidelity varies with dimensionality and directionality.
	\par The \textit{root node} is the foremost node of the tree, and the nodes without any child nodes are defined as the \textit{leaf nodes}. The \textit{depth} (or \textit{height}) $d$ of a node in a binary tree is defined as the total number of edges from the root node to the target node. 
	The number of nodes of a full binary tree is $2d+1 \leq N \leq 2^{d+1}-1$. We consider $N$ as a function of depth $d$.
	\par
\begin{figure*}[htbp]
		\centering
		
		\begin{minipage}[t]{0.45\textwidth}
			\centering
			\includegraphics[scale = 0.18]{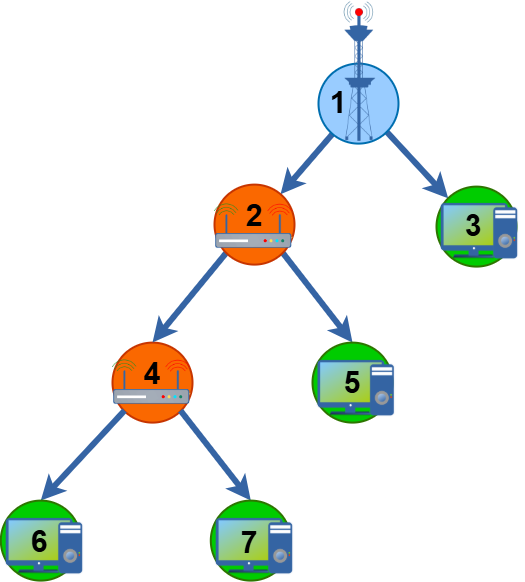}
			\vspace{1mm}
			\par(a) DABT
			\label{fig:da}
		\end{minipage}
		\hfill
		%
		\begin{minipage}[t]{0.45\textwidth}
			\centering
			\includegraphics[scale = 0.18]{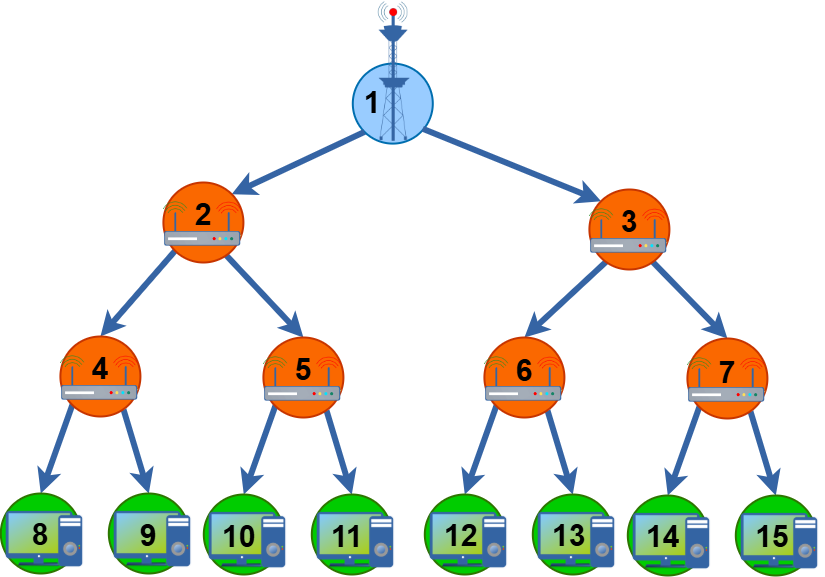}
			\vspace{1mm}
			\par (b) DSBT
			\label{fig:ds}
		\end{minipage}
		\hfill \\
		%
		\begin{minipage}[t]{0.45\textwidth}
			\centering
			\includegraphics[scale = 0.18]{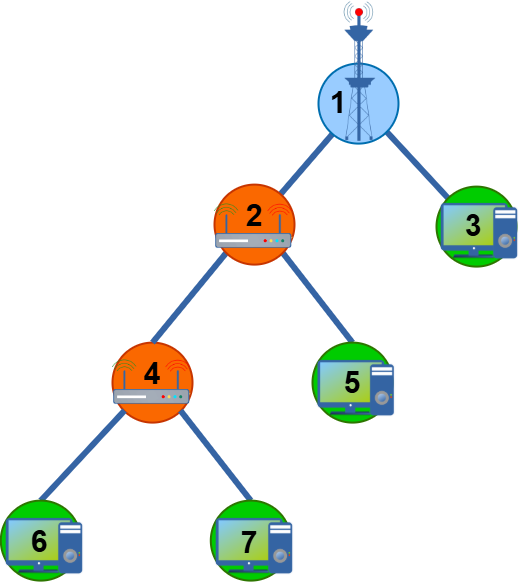}
			\vspace{1mm}
			\par (c) UABT
			\label{fig:ua}
		\end{minipage}
		\hfill
		%
		\begin{minipage}[t]{0.45\textwidth}
			\centering
			\includegraphics[scale = 0.18]{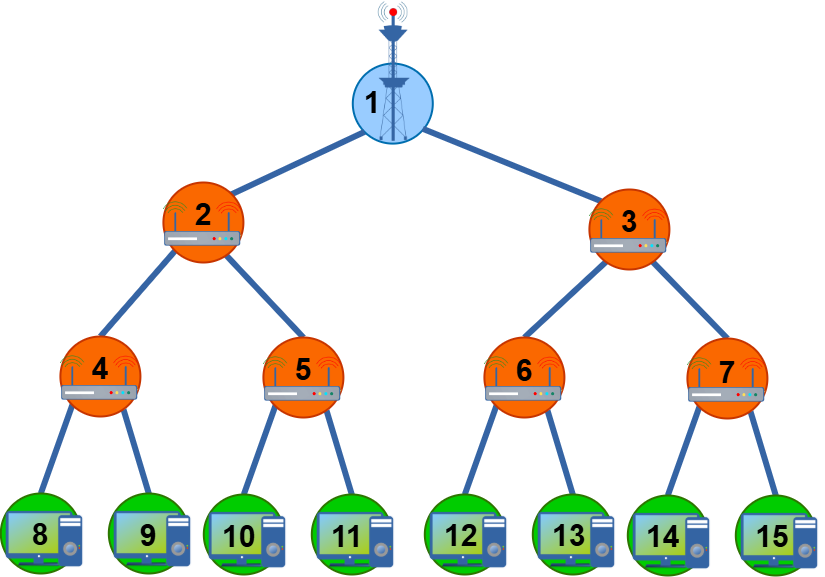}
			\vspace{1mm}
			\par(d) USBT
			\label{fig:us}
		\end{minipage}
		\hfill
		
		\caption{\textbf{Top Panel:} We have (a) DABT and (b) DSBT with 7 and 15 nodes, respectively. \textbf{Bottom Panel:} We have (c) UABT and (d) USBT with 7 and 15 nodes, respectively. The depth of each binary tree is $d$ = 3.}
		\label{fig:bt}
	\end{figure*}

\begin{figure*}[htbp]
		\centering
		\includegraphics[scale=0.13]{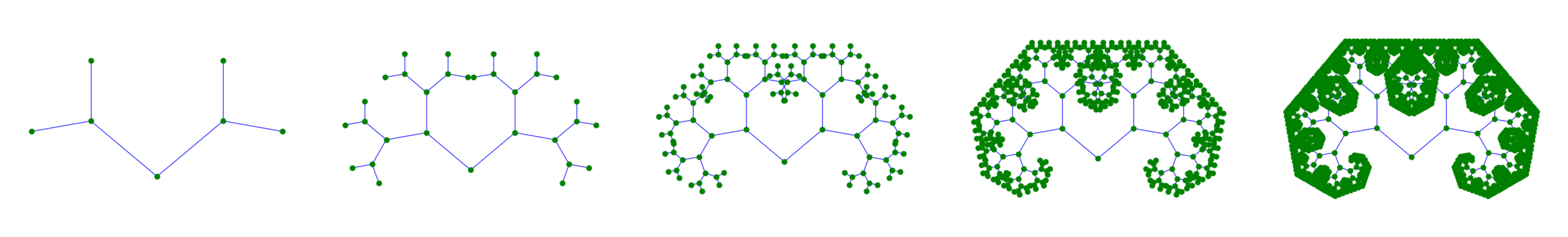}
		\caption{The fractal tree from an undirected symmetric binary tree. From left to right, the depths of the binary trees are 2, 4, 6, 8, and 10, respectively.}
		\label{fig:ft}
	\end{figure*}

 In Figs.\ \ref{fig:bt}(a), \ref{fig:bt}(b), \ref{fig:bt}(c), and \ref{fig:bt}(d), we showed four different cases of binary tree network \textit{i.e.} directed asymmetric binary tree (DABT), directed symmetric binary tree (DSBT), undirected asymmetric binary tree (UABT), and undirected symmetric binary tree (USBT) respectively. The root node (depicted in skyblue colour) of the tree is node number 1 in each of the cases. In the case of the asymmetric trees (DABT and UABT), the leaf nodes (depicted in green colour) are 3, 5, 6, and 7, whereas the leaf nodes are 8, 9, 10, 11, 12, 13, 141 and 15 in the case of symmetric trees (DSBT and USBT).  The depth of each tree is $d=3$ in Figs.\ \ref{fig:bt}(a), (b), (c), and (d). The total number of nodes in DABT, DSBT, UABT and USBT is 7, 15, 7, and 15, respectively.
	Note that a fractal tree (FT) can be created by splitting one branch into two sub-branches, which is symmetric, and doing this step infinitely. The fractal trees are studied in the context of telecommunications as well \cite{chen2020capacityfractald2dsocial, Ezhumalai2021}. This tree is similar to the case of the undirected symmetric binary tree. As depicted in Fig. \ref{fig:ft}, if we constantly increase the depth of the undirected symmetric binary tree, then it will lead to a fractal tree when $d$ is very large.
	%
	%
	%
	%
			\section{Average  Teleportation Fidelity in Binary Tree Network} \label{sec3}
			In this section, we consider a repeater-based entangled network of $N$ nodes arranged in a binary tree structure. All the edges of our network are made up of Werner states given in Eq.\ \eqref{eq:wener} parameterized by the parameter $p$. In this section, we consider four cases of the binary tree network: 1) directed asymmetric binary tree (DABT), 2) directed symmetric binary tree (DSBT), 3) undirected asymmetric binary tree (UABT), 4) undirected symmetric binary tree (USBT) and obtain the analytical expression for average teleportation fidelity for each of these networks. In practical cases, an intermediate link will be destroyed after a swapping process. Therefore, communication between the source and target pair is impossible if there is only one Werner state present between any two nodes. Therefore, there is an ensemble of Werner states between any source-target pair. Due to this fact, the qubit can be transferred from any source and target pair. This is the reason we considered all possible paths in the binary tree network.
			\begin{figure}[h]
				\centering
				\includegraphics[width=0.8\linewidth]{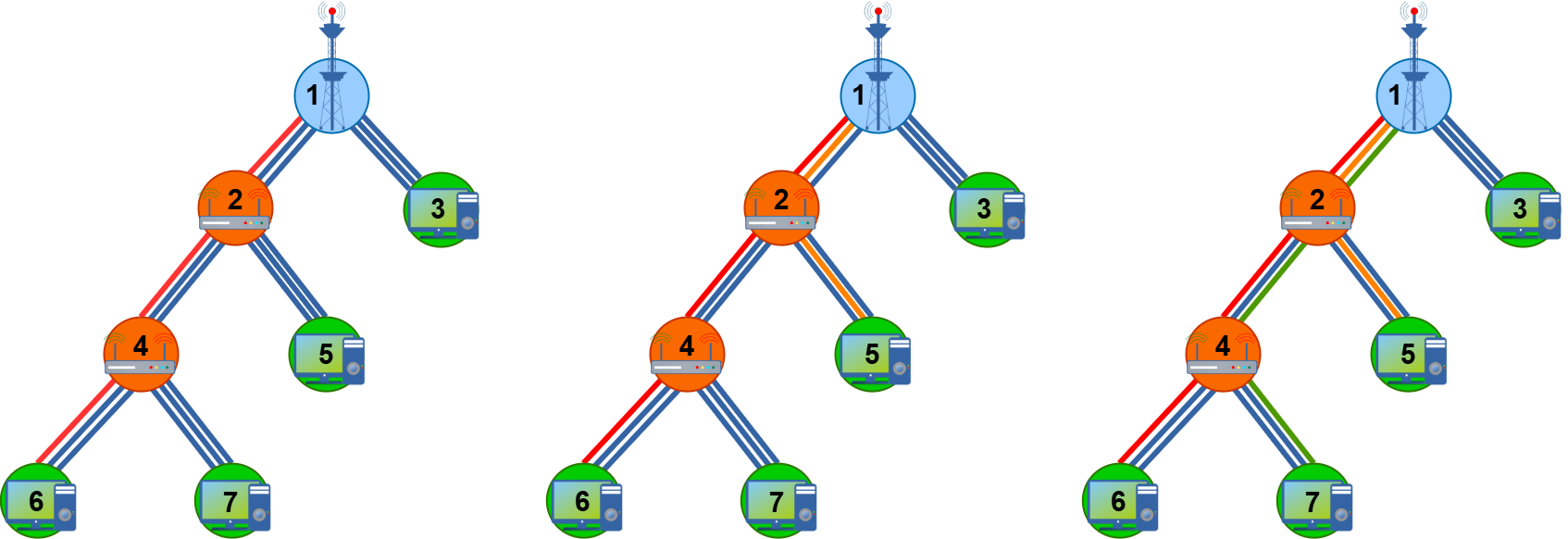}
				\caption{Toy Example: In this UABT network with 7 nodes, the first connection is established between 1 and 6 (shown using the red line). After that, the connection is established between 1 and 5 (orange line) and finally between 1 and 7 (green line).}
				\label{fig:multiplexing}
			\end{figure}
			%
			%
			%
			%
			
			\subsection{Notations}
			The related and useful notation is shown in the Table \ref{tab:notation} . We have used the same notations through out the article until stated otherwise. 
			\begin{table}[htbp]
				\centering
				\begin{tabular}{|c|l|}
					\hline
					\textbf{Notation} & \textbf{Explanation} \\
					\hline
					$p$ & Werner state parameter\\
					\hline
					$d$ & The depth of the binary tree  \\
					\hline
					$N$ & The number of nodes in the binary tree \\
					\hline
					$l_d^{(i)}$ & the number of $i$ length path at level $d$\\
					\hline
					$\mathcal{F}_i$ & Teleportation fidelity of $i$ length path\\
					\hline
					$F^{\text{tel}}_{\text{avg}}$ & Average of maximum teleportation fidelity\\
					\hline
				\end{tabular}
				\caption{Notations}
				\label{tab:notation}
			\end{table}

			\subsection{Directed Binary Tree}
			In this subsection, we consider two types of directed binary trees. These include both asymmetric and symmetric cases. Here, the directed link indicates that teleportation is permitted only in one direction. As an example, in Fig. \ref{fig:bt}(a) and (b), a directed link from node 1 to node 6 implies that teleportation can occur from node 1 to node 6, but not from node 6 to node 1, and so on.\\
			
			\noindent \textbf{Directed Asymmetric Binary Trees (DABT) :} Here we consider a directed and asymmetric binary tree where we have thought that each node represents a Werner state as shown in Fig.\ \ref{fig:bt}(a). Let us denote the number of levels as $d$ of the binary tree. Therefore, $F^{\text{tel}}_{\text{avg}}$ in this case can be expressed as
			\begin{eqnarray}
				\left.F^{\text{tel}}_{\text{avg}} (d,p)\right|_{\text{DABT}} = \frac{\sum\limits_{i=1}^{d} [2+2(d-i)]\mathcal {F}_i}{\sum\limits_{i=1}^{d} [2+2(d-i)]} = \frac{1}{2}+\frac{p[p(p^d-1)+d(1-p)]}{d(d+1)(1-p)^2}.
				\label{f_dabt}
			\end{eqnarray}
			where $\mathcal{F}_i = \frac{1}{2}(1+p^i)$. For further analysis, we are expressing $F^{\text{tel}}_{\text{avg}}$ in terms of total number of node ($N$) where $N=2d+1$. In that case, we have
			\begin{eqnarray}
				\left.F^{\text{tel}}_{\text{avg}} (N,p)\right|_{\text{DABT}} = \frac{\sum\limits_{i=1}^{(N-1)/2} (N+1-2i)\mathcal {F}_i}{\sum\limits_{i=1}^{(N-1)/2} (N+1-2i)} = \frac{1}{2}+\frac{2p[2p(p^{\frac{N-1}{2}}-1)+(N-1)(1-p)]}{(N-1)(N+1)(1-p)^2}.
				\label{f_dabtn}
			\end{eqnarray}
			The derivation of the formula is given in appendix \ref{app:A}. \\ 
			%
			%
			%
			
			%
			\noindent \textbf{Directed Symmetric Binary Trees (DSBT) :}
			Here we consider a directed and symmetric binary tree where we have thought that each node represents a Werner state, and it is shown in Fig.\ \ref{fig:bt}(b). In this case, $F^{\text{tel}}_{\text{avg}}$ can be expressed as
			\begin{eqnarray}
				\left.F^{\text{tel}}_{\text{avg}} (d,p)\right|_{\text{DSBT}} = \frac{\sum\limits_{i=1}^{d} [2^{d+1} - 2^i]\mathcal{F}_i}{\sum\limits_{i=1}^{d}[2^{d+1} - 2^i]}
				=
				\begin{cases}
					\displaystyle
					\frac{1}{2} +\frac{2^{d+1} - d - 2}{2\left[(d - 1) \cdot 2^{d+1} + 2\right]} , & \text{if }~ p = \frac{1}{2}, \\[12pt]
					\displaystyle
					\frac{1}{2} +\frac{p\left[(2^d - 1) - p(2^{d+1}-1) + 2^d p^{d+1}\right]}{2(1-p)(1-2p)\left[(d - 1)\cdot 2^d + 1\right]}, & \text{otherwise},
				\end{cases}
				\label{f_dsbt}
			\end{eqnarray}
			where $d$ is the number of levels, like the previous case and $\mathcal{F}_i = \frac{1}{2}(1+p^i)$. It is expressed $F^{\text{tel}}_{\text{avg}}$ in terms of the total number of nodes ($N$) where $N=2^{d+1}-1$, for additional analysis. Hence,
			%
			\begin{eqnarray}
				\left.F^{\text{tel}}_{\text{avg}} (N,p)\right|_{\text{DSBT}} &=& \frac{\sum\limits_{i=1}^{\log_2[(N+1)/2]} (N+1-2^i)\mathcal{F}_i}{\sum\limits_{i=1}^{\log_2[(N+1)/2]}(N+1-2^i)} \\
				&=& \begin{cases}
					\displaystyle
					\frac{1}{2} +\frac{N-\log_2(N+1)}{2\left[(N+1)\log_2(N+1)-2N \right]} , & \text{if }~ p = \frac{1}{2}, \\[12pt]
					\displaystyle
					\frac{1}{2}+\frac{p[N(1-2p)+(N+1)p^{\log_2(N+1)}-1]}{4(1-p)(1-2p)\left[(N+1)\log_2(N+1)-2N \right]}, & \text{otherwise}.
				\end{cases}
				\label{f_dsbtn}
			\end{eqnarray}
			%
			A detailed derivation of the formula is provided in appendix \ref{app:A}. For both the cases of directed binary trees, $p=1$ leads to $F^{\text{tel}}_{\text{avg}} (N,p) = 1$.
			
			
			\subsection{Undirected  Binary Trees }
			In this subsection, we consider two types of undirected binary trees, which incorporate both asymmetric and symmetric cases. Here, the undirected link indicates bidirectional teleportation. As an example, in Figs.\ \ref{fig:bt}(c) and \ref{fig:bt}(d), teleportation can occur from node 1 to node 6 as well as from node 6 to node 1,  and so on \textit{i.e.} no restriction due to the directionality.\\
			
			\noindent \textbf{Undirected  Asymmetric Binary Trees (UABT) :} Here we consider an undirected asymmetric binary tree network (UABT) where each link represents a Werner state, where each Werner state is parameterized by the classical mixing parameter $p$. It is shown in Fig.\ \ref{fig:bt}(c). In the UABT network, we start with $3$ nodes at level $h=1$, and after that we add two nodes at every level. So in that case, $N$ will be odd numbers starting from $3$ (\textit{i.e.} $N=3, 5, 7 \cdots$).
			\par 
			In this context, the average network fidelity for teleportation $\left.F^{\text{tel}}_{\text{avg}} (d,p)\right|_{\text{UABT}}$ can be obtained as
			%
			\begin{eqnarray}
				\left.F^{\text{tel}}_{\text{avg}} (d,p)\right|_{\text{UABT}} &=&
				\frac{2d \times \mathcal F_1+(1+3(d-1))\mathcal F_2+\sum\limits_{i=3}^{d+1} (2+4(d-i+1))\mathcal F_i}{2d+(1+3(d-1))+\sum\limits_{i=3}^{d+1}(2+4(d-i+1))}\nonumber \\
				&=& \frac{1}{2}+\frac{2dp - (d + 2)p^2 - 2p^3 - dp^4 + 2p^{d+2} + 2p^{d+3}}{2d(2d + 1)(1 - p)^2},
				\label{f_uabt}
			\end{eqnarray}
			%
			where $\mathcal F_i \equiv (1+p^i)/2$. The derivation of the formula is given in the appendix. Now, let us change the $d$ with $N$, where $N = 2d+1$, gives us,
			\begin{align}
				\left.F^{\text{tel}}_{\text{avg}} (N,p)\right|_{\text{UABT}} 
				&= \dfrac{ (N-1)\mathcal{F}_1 + \left(1 + \frac{3}{2}(N - 3)\right)\mathcal{F}_2 + \sum\limits_{i=3}^{(N+1)/2} \left(2 + 4\left(\frac{N+1}{2} - i\right)\right)\mathcal{F}_i}{(N-1) + \left(1 + \frac{3}{2}(N - 3)\right) + \sum\limits_{i=3}^{(N+1)/2} \left(2 + 4\left(\frac{N+1}{2} - i\right)\right)} \nonumber \\
				&= \frac{1}{2} + \frac{2p(N - 1) - (N + 3)p^2 - 4p^3 - (N - 1)p^4 + 4p^{\frac{N+3}{2}} + 4p^{\frac{N+5}{2}}}{2N(N - 1)(1 - p)^2}.
				\label{f_uabtn}
			\end{align}
			
			The full derivation of the formula is given in appendix \ref{app:B}. \\
			
			
			%
			%
			%
			\noindent \textbf{Undirected  Symmetric Binary Trees (USBT)}: Let us consider an undirected symmetric binary tree network (USBT) as shown in Fig.\ \ref{fig:bt}(d). We also consider each link representing a Werner state, parameterized by $p$. In this case, $F^{\text{tel}}_{\text{avg}}$ can be written as 
			\begin{eqnarray}
				\left.F^{\text{tel}}_{\text{avg}} (d,p)\right|_{\text{USBT}} =
				\frac{\sum\limits_{i=1}^{2d} l_d^{(i)} \mathcal F_i}{\sum\limits_{i=1}^{2d} l_d^{(i)}},
				\label{f_usbt}
			\end{eqnarray}
			where $\mathcal F_i \equiv (1+p^i)/2$ and $l_d^{(i)}$ is the number of $i$ length path at level $d$ which is given by 
			
			\begin{equation*}
				l_d^{(i)}=
				\begin{cases}
					3 \cdot 2^{d + \frac{i}{2} - 1} - (2d-i+5) \cdot 2^{i - 2},& \text{if} ~d<i\leq 2d ~ \text{and} ~i~ \text{is even}\\
					2^{d + \frac{i+1}{2}} - (2d-i+5) \cdot 2^{i - 2},& \text{if} ~d<i\leq 2d ~\text{and} ~i ~\text{is odd}\\
					3 \cdot 2^{d + \frac{i}{2} - 1} - (i+3) \cdot 2^{i - 2},& \text{if} ~i\leq d ~\text{and} ~i ~\text{is even}\\
					2^{d + \frac{i+1}{2}} - (i+3) \cdot 2^{i - 2}, & \text{if} ~i\leq d ~ \text{and} ~ i ~ \text{is odd}\\
					0,& \text{otherwise}.
				\end{cases}
			\end{equation*}
			%
			\noindent Like the previous cases, here we can also express Eq.\ \eqref{f_usbt} in terms of $N$ where $N = 2^{d+1}-1$. The full derivation of the formula is given in appendix \ref{app:B}. Similarly, for the both undirected binary trees, $F^{\text{tel}}_{\text{avg}} (N,p) = 1$ when $p=1$. {In the case of FT, the same formula holds.}
			
			
			\section{Quantum Advantage in Binary Tree Networks} \label{sec4}
			In this subsection, we explore the idea of quantum advantage in a network which is given by the simple condition $F^{\text{tel}}_{\text{avg}} >2/3$. This is because without an entangled state, each path can achieve a maximum teleportation fidelity of 2/3. Hence, the average teleportation fidelity will achieve a maximum of 2/3 in the absence of entangled states. \\
			
\begin{figure}[h]
				\centering
				\begin{minipage}[b]{0.45\linewidth}
					\centering
					\includegraphics[scale=0.25]{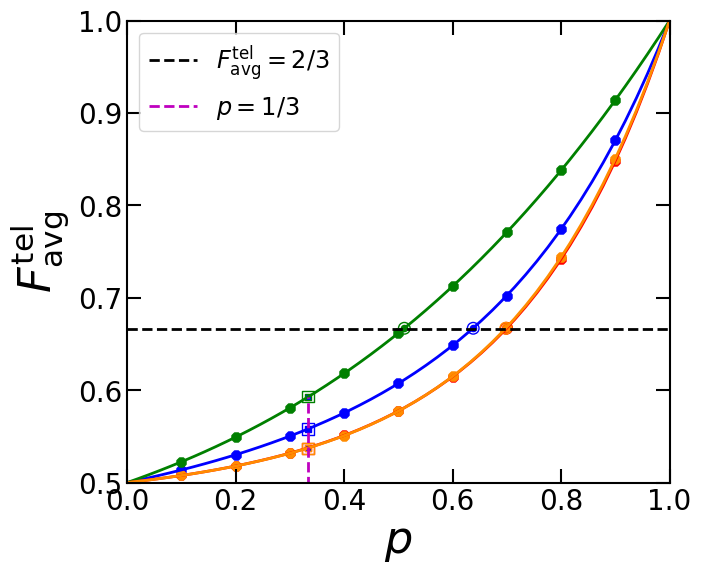}\\ 
					\quad (a)
				\end{minipage}
				\begin{minipage}[b]{0.45\linewidth}
					\centering
					\includegraphics[scale=0.25]{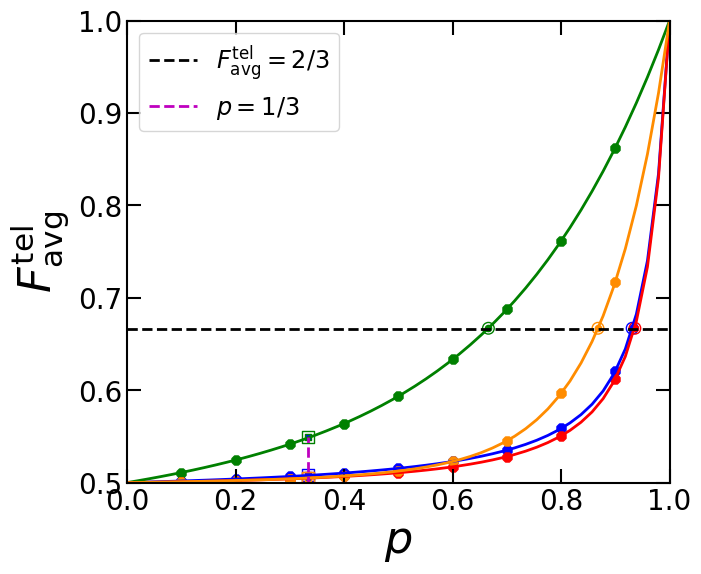}\\
					\quad (b)
				\end{minipage}
				\caption{The plot of $F^{\text{tel}}_{\text{avg}}$ vs for different values of $p$ in case of all four types of binary trees (a) for $N = 15$ (In case of DABT and UABT, $d = 7$ and for DSBT and USBT, $d = 3$), (b) $N=127$. (In case of DABT and UABT, $d = 63$ and for DSBT and USBT, $d = 6$).  Here, the blue, green, red and orange lines denote DABT, DSBT, UABT and USBT, respectively.}
				\label{Quant Ad}
			\end{figure}

			\noindent \textbf{Quantum Advantage:} 
			For instance, quantum advantage was explained in reference \cite{mylavarapu2024teleportation} by the condition $F^{\text{tel}}_{\text{avg}}>2/3$. We know that, Werner states as a resource, shared between two parties is helpful for teleportation when the input parameter $p>1/3$. 
			Nevertheless, in a network situation, $p>1/3$ for each link is insufficient to generate an average fidelity more than $2/3$. Thus, for each one of these networks, we calculate $F^{\text{tel}}_{\text{avg}}$ at $p=1/3$ as shown in Fig.\ \ref{Quant Ad}(a) ($N=15$) and in Fig.\ \ref{Quant Ad}(b) for $N=127$.
			In case of $N=15$, $F^{\text{tel}}_{\text{avg}}$ for DABT, DSBT, UABT, and USBT, respectively, is obtained as $0.558, 0.593, 0.537$, and $0.537$ (Fig.\ \ref{Quant Ad}(a)). They are indicated in the diagram by double square markers (see the dashed vertical line). In a large network (Fig.\ \ref{Quant Ad}(b), $N=127$) the values of $F^{\text{tel}}_{\text{avg}}$ for the same series of networks are shown in the diagram as double square markers (along dashed vertical line) and are 0.508, 0.549, 0.505, and 0.505, respectively.
			\par 
			There are various kinds of comprehension for the concept of quantum advantage in the distributed teleportation ( \textit{i.e.} teleportation between all possible sources and targets) of a repeater network. The $F^{\text{tel}}_{\text{avg}}$ has been plotted with the input state parameter in Figs.\ \ref{Quant Ad}(a) and \ref{Quant Ad}(b) to determine the value of $p$ at which each of these four networks exhibits quantum advantages. The quantum advantage is demonstrated by taking into account all types of trees that have the same number of nodes, meaning that depth $d$ differs for symmetric and asymmetric trees. We presented the results for $N=15$ in Fig.\ \ref{Quant Ad}(a) and for $N=127$ in Fig.\ \ref{Quant Ad}(b).  For all values of $p$ beyond $0.638$, $0.511$, $0.699$, and $0.698$, respectively, DABT, DSBT, UABT, and USBT exhibit quantum advantage shown in Fig.\ \ref{Quant Ad}(a). The values are indicated in the diagram by double-circle markers (see the dashed horizontal line). 
			The quantum advantage is also obtained for a range of $p$ larger than $0.931$, $0.666$, $0.936$, and $0.868$, respectively, for $N=127$ (Fig.\ \ref{Quant Ad}(b), dashed horizontal line).

			

			For $N=15$, given that directed binary trees include fewer pathways than undirected binary trees, it is intriguing to note that they consistently outperform the latter in terms of permitting a larger range of $p$.  
			The average pathlength $\langle L \rangle$ has values of $3$, $1.83$, $3.45$, and $3.5$ for DABT, DSBT, UABT, and USBT, respectively, for $N=15$. 
			However, for DABT, DSBT, UABT, and USBT, the corresponding values of $\langle L \rangle$ are $21.67$, $3.19$, $22.31$, and $8.35$ in the case of $N=127$. In both circumstances, the DSBT value of $\langle L \rangle$ is at its minimum. 
			It can be shown that for large $N$,  $\langle L \rangle$, is minimum, compared to the other three networks. 
			These observations are consistent with the $F^{\text{tel}}_{\text{avg}}$ values of all the networks.
			
\begin{figure}[h]
				\centering
				\includegraphics[scale=0.25]{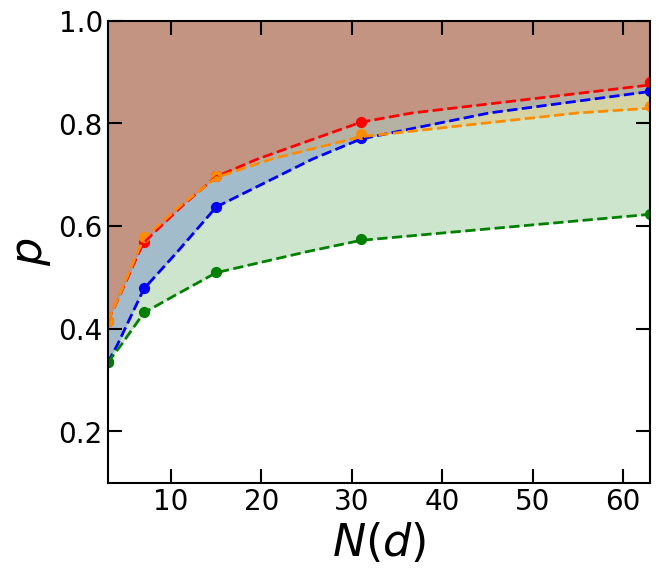}
				\caption{ The contour plot of $p$ vs $N(d)$ when $F^{\text{tel}}_{\text{avg}} = 2/3$.  Here, the blue, green, red and orange lines denote DABT, DSBT, UABT and USBT, respectively. The corresponding shaded region belongs to $F^{\text{tel}}_{\text{avg}} > 2/3$.}
				\label{Quant Ad1}
			\end{figure}

			
			In Fig.\ \ref{Quant Ad1}, we have also plotted a contour diagram to demonstrate how the threshold value of $p$ changes with the increasing $N(d)$ for all these four networks. Here we keep the $N$ fixed for all the cases, and hence $d$ is not constant for the asymmetric and symmetric cases. It can be seen that DSBT performs better than the other networks on average, making it more suitable as a teleportation network.

\begin{figure}[!h]
				\centering
				\begin{minipage}[b]{0.24\linewidth}
					\centering
					\includegraphics[scale=0.22]{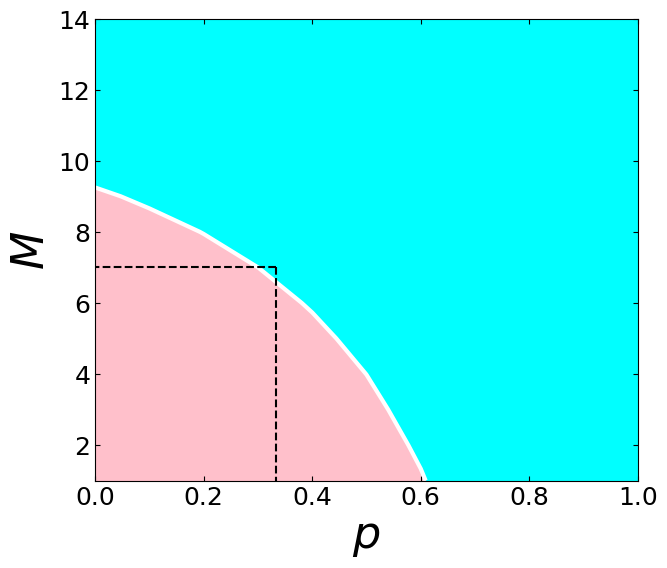}\\
					\qquad (a)
				\end{minipage}
				\hfill
				\begin{minipage}[b]{0.24\linewidth}
					\centering
					\includegraphics[scale=0.22]{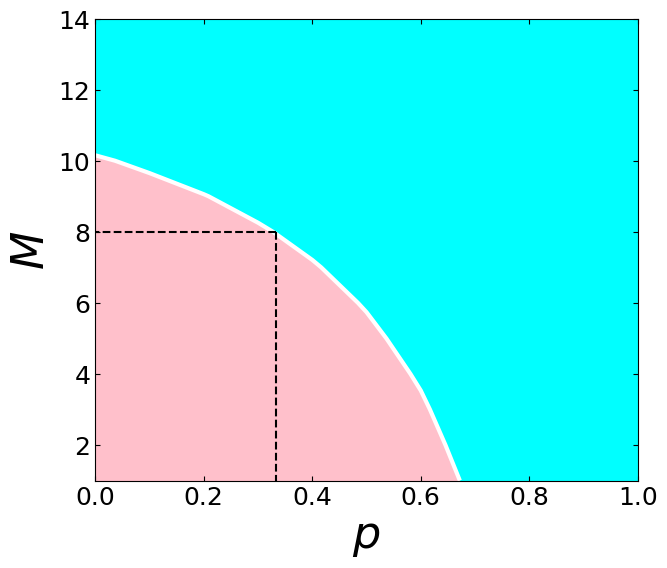}\\
					\qquad (b)
				\end{minipage}
				\hfill
				\begin{minipage}[b]{0.24\linewidth}
					\centering
					\includegraphics[scale=0.22]{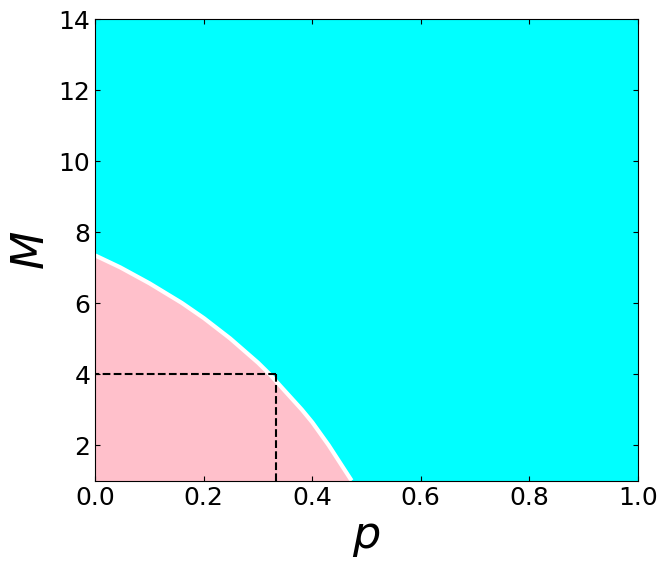}\\
					\qquad (c)
				\end{minipage}
				\hfill
				\begin{minipage}[b]{0.24\linewidth}
					\centering
					\includegraphics[scale=0.22]{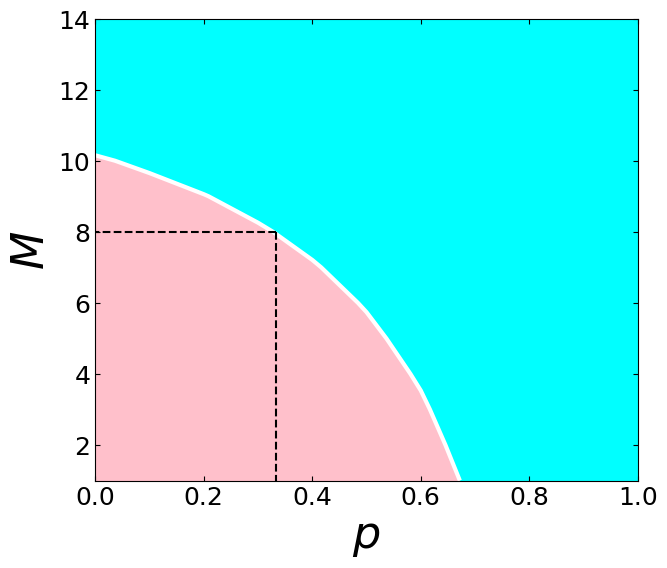}\\
					\qquad (d)
				\end{minipage}
				
				\caption{The contour plot of $F^{\text{tel}}_{\text{avg}}$: (a) DABT, (b) UABT, (c) DSBT, and (d) USBT.  In the case of $p=0.333$, the network achieves $F^{\text{tel}}_{\text{avg}}=2/3$ when $M = 7$ and 8, 4 and 8 for DABT, UABT, DSBT, USBT, respectively.}
				
				\label{contour}
			\end{figure}

			\noindent \textbf{Maximally Entangled links :} Next, we look into a scenario where we consider some link to be a maximally entangled state. In reference \cite{mylavarapu2024teleportation}, the role of maximally entangled states in determining the average fidelities of loopless topologies like start, chain and branching trees. The assignment of maximally entangled states is also going to play a crucial role in determining the average fidelity. In Fig.\ \ref{contour}(a), \ref{contour}(b), \ref{contour}(c), and \ref{contour}(d), we have used contour plot to show how $F^{\text{tel}}_{\text{avg}}$ changes as a function of the number of maximally entangled states ($M$) vs $p$. The pink region in the figures indicate the region where $F^{\text{tel}}_{\text{avg}}<\frac{2}{3}$. The cyan region in the figures indicates the ranges of $M$ and $p$ for which the network teleportation fidelity is  $F^{\text{tel}}_{\text{avg}}>\frac{2}{3}$. It is clear from each graph that for a lower value of $p$, we require a larger number of maximally entangled states in each of these graphs for $F^{\text{tel}}_{\text{avg}}$ to cross the threshold limit $2/3$. As expected for a larger value of $p$, we do not require maximally entangled states for $F^{\text{tel}}_{\text{avg}}$ to cross the threshold limit. For $N=15$, the threshold value of $M$ required  for quantum advantage are $4$, $7$, $8$, and $8$ for DSBT, DABT, UABT, and USBT. Among the four graphs, DSBT performs the best as there is a larger range of $p$ for which there is a quantum advantage in the network.
			
				

\begin{figure}[!h]
				\centering
				\begin{minipage}[b]{0.24\linewidth}
					\centering
					\includegraphics[scale=0.18]{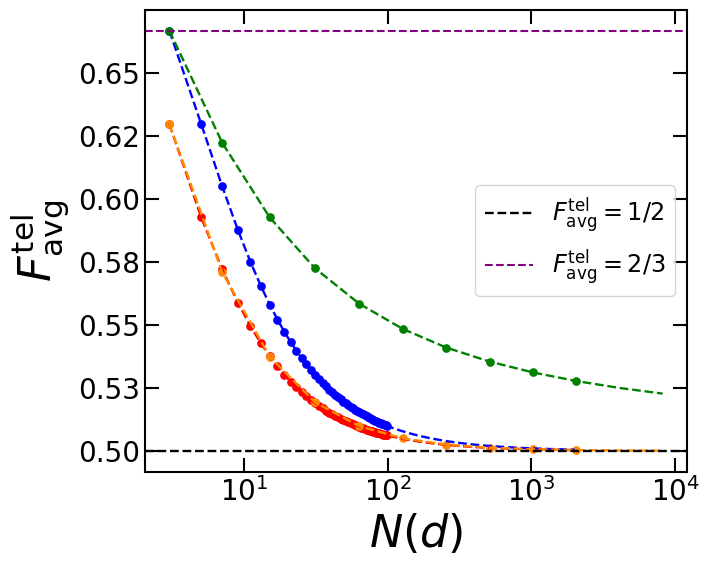}\\
					\qquad (a)
				\end{minipage}
				\hfill
				\begin{minipage}[b]{0.24\linewidth}
					\centering
					\includegraphics[scale=0.18]{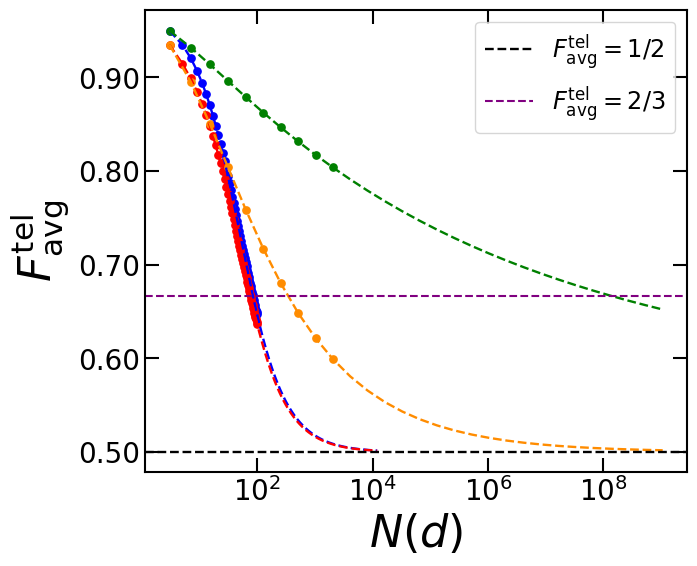}\\
					\qquad(b)
				\end{minipage}
				\hfill
				\begin{minipage}[b]{0.24\linewidth}
					\centering
					\includegraphics[scale=0.18]{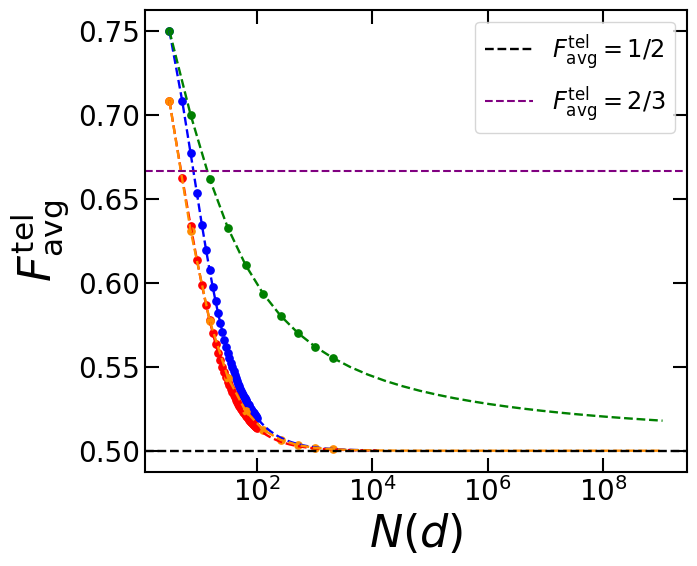}\\
					\qquad(c)
				\end{minipage}
				\hfill
				\begin{minipage}[b]{0.24\linewidth}
					\centering
					\includegraphics[scale=0.18]{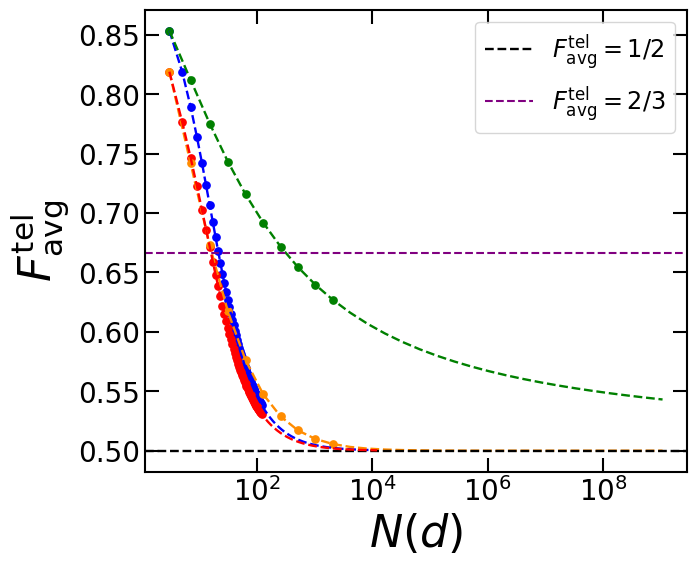}\\
					\qquad (d)
				\end{minipage}
				\caption{ The plot of $F^{\text{tel}}_{\text{avg}}$ vs $N(d)$ in the case of all four types of binary trees (a) for $p=0.333$, (b) for $p=0.9$, (c) $p=1/2$, (d) $p = 1/\sqrt{2}$. Here, the blue, green, red and orange lines denote DABT, DSBT, UABT and USBT, respectively. For all the cases, $F^{\text{tel}}_{\text{avg}}$ converges to $1/2$ when $N$ is large.}
				\label{large N 1}
			\end{figure}

			%
			%
			%
				\noindent \textbf{Large $N$ limit:} 
				For DABT, we have analytically and numerically demonstrated $F^{\text{tel}}_{\text{avg}}$  reaches $1/2$ for a large $d$ (large $N$). In Eq.\ \eqref{DABTpluserror}, and the Fig.\ \ref{large N error} of appendix \ref{app:B}, the additional term $\varepsilon$ is plotted as a function of $N(d)$ that decreases to zero as the number of nodes in the tree increases.  This is also true for UABT, USBT, and DSBT (see the Eqns.\ \eqref{UABTpluserror}, \eqref{DSBTpluserror}, and Fig.\ \ref{large N error} in the appendix). Clearly, regardless of the network configuration, the plot (in the appendix \ref{app:B}) indicates that additional parts added with $1/2$ converge to zero for a large $N(d)$ limit. In case of DSBT, the additional part $\varepsilon(N) = \mathcal{O}\lrfb{1/\log_2 N}$ (See appendix \ref{app:A}) and thus it decays slowly with $N(d)$ compared to other binary trees. Hence, we get a larger additional term in the case of DSBT.
				For numerical verification, we computed the $F^{\text{tel}}_{\text{avg}}$ with increasing $N$ for various $p$ values, namely $p=0.333$ and $p=0.9$, in order to see the quantum advantage in the large $N$ limit. Alongside this, we have also plotted for $p=1/2$ and $1/\sqrt{2}$ to verify the formula with the numerical result. The results are reported in 
				Figs.\  \ref{large N 1}(a)($p=0.333$), \ref{large N 1}(b)($p=0.9$), \ref{large N 1}(c)($p=1/2$), and \ref{large N 1}(d)($p=1/\sqrt{2}$). We observe $F^{\text{tel}}_{\text{avg}}$ is converging to $1/2$ but with different rates. 
				It is evident that for low values of $p$ ($p=0.333$), each of these networks $F^{\text{tel}}_{\text{avg}}$ rapidly stabilizes to the value $1/2$. 
				Conversely, for very big $N$ ($\sim 10^8$), DSBT exhibits quantum advantage at $p=0.9$. Nevertheless, inside the $10^3$ network size, DABT, UABT, and USBT lose the quantum advantage. Note that we aim to display comparison results from four networks here. For this, the depth $d$ of each graph is selected so that every network has the same size. { Moreover, in case of large $N$, the USBT turns into a FT and hence $F^{\text{tel}}_{\text{avg}}$ converges to the value $1/2$ in case of a FT. As depicted, it will not retain quantum advantage up to very large $N$ compared to DSBT.}

			%
			

		\section{Average teleportation fidelity for uniform distribution} \label{sec5}
\begin{figure}[h]
			\centering
			\includegraphics[scale = 0.25]{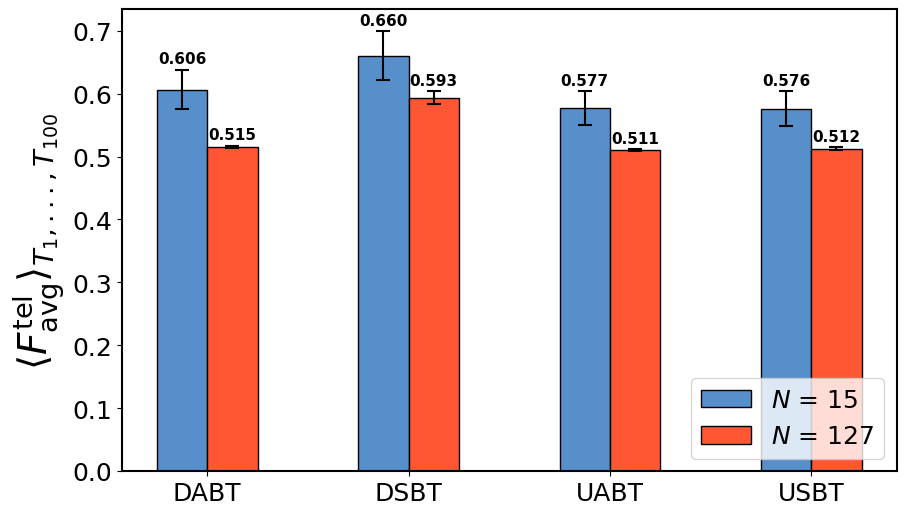}
			\caption{The sample average of Teleportation fidelity $\langle F_{\text{avg}}^{\text{tel}}\rangle _{T_1,...,T_{100}}$ for four different binary trees for $N=15$ and $127$ with $p$ values are chosen from uniform distribution. There are $100$ iterations of $p$ values. The vertical black bars show the standard deviation of the simulation.}
			\label{uniform1}
		\end{figure}

		In a real network (based on photonic qubits), parameters like detector efficiency, the likelihood of a photon being emitted in the correct mode, the probability of a photon being successfully sent into the fiber, attenuation losses in the fiber, etc., cause the qubit's coherence to be lost. This loss also varies from node to node, as it depends upon the distance between the nodes and also on the individual capacity of repeaters. This variation can be modeled by different Werner state parameters $p_i$ for each link \cite{Ferreira2024}. Accordingly, the probability that a single transmission attempt succeeds can be written as
		\begin{eqnarray}
			p_{\mathrm{noise}} = p_{\mathrm{det}} \times 10^{-\alpha d/10},
		\end{eqnarray}
		where $\alpha$ is the fibre attenuation coefficient and $d$ is the transmission distance. The time required to have one attempt to transfer a qubit is $\tau = d/c$ with $c$ is the speed of light in the fibre. If the photon is successfully detected, then $p_{\mathrm{det}} = 1$. In this scenario, $F^{\text{tel}}_{\text{avg}}$ goes down if we increase the intra-node distance ($d$) \cite{mylavarapu2024teleportation}. For a fixed number of nodes $N$ in each class of binary trees, we assign weights $\{p_i\}$ to the edges, where each weight is independently drawn from a uniform distribution, $p_i \sim \mathcal{U}(0,1)$. Since a binary tree with $N$ nodes contains $N - 1$ edges, each tree configuration is associated with $N - 1$ such weights. To include statistical robustness, we repeat this random weight assignment process 100 times ($T_1,T_2,...,T_{100}$) for each tree instance, and each time, the corresponding average teleportation fidelity, $F^{\text{tel}}$. We then take the average of these values across all trials to obtain $\langle F_{\text{avg}}^{\text{tel}}\rangle _{T_1,...,T_{100}}$. The result is shown as a histogram in Fig.\ \ref{uniform1}.
		\par For a large number of trials, the sample average of average teleportation fidelity will be almost equal to the average teleportation fidelity for the situation when all the weights $p_i = 1/2$. This is due to the fact that the expectation value of $p_i$ is $1/2$ when we consider $\mathcal{U}(0,1)$. In this setting, $F_{\text{avg}}^{\text{tel}}$ can be computed $F_{\text{avg}}^{\text{tel}}$ as
		{\begin{small}
				\begin{eqnarray}
					F_{\text{avg}}^{\text{tel}}(N,p) = \frac{\sum_{i}\text{Total Number of $i$ length Paths}\times \mathcal{F}_i}{\text{Total Number of Paths}},
				\end{eqnarray}
		\end{small}}
		\noindent \\ where $\mathcal{F}_i = \frac{1}{2}(1+p_1p_2\hdots p_i)$ and $p_1 = p_2 = \hdots = p_i = 1/2$. It can be seen that for very large number of trials (In this case $T=100$), $\langle F_{\text{avg}}^{\text{tel}}\rangle _{T_1,...,T_{100}} \to F_{\text{avg}}^{\text{tel}} (p=1/2)$. To such illustrations, for $N=15$ and 127 are shown in Table \ref{tab:uniform2}.
		\begin{table}[h]
			\centering
			\resizebox{\textwidth}{!}{
				\setlength{\tabcolsep}{7pt}
				\begin{tabular}{|c|c|c|c|c|c|c|c|c|}
					\hline
					& \multicolumn{2}{|c|}{\textbf{DABT}} & \multicolumn{2}{|c|}{\textbf{DSBT}} & \multicolumn{2}{|c|}{\textbf{UABT}} & \multicolumn{2}{|c|}{\textbf{USBT}} \\
					\hline
					\textbf{Nodes} & \textbf{$F_{\text{avg}}^{\text{tel}}$} & \textbf{$\langle F_{\text{avg}}^{\text{tel}}\rangle _{T_1,...,T_{100}}$} & \textbf{$F_{\text{avg}}^{\text{tel}}$} & \textbf{$\langle F_{\text{avg}}^{\text{tel}}\rangle _{T_1,...,T_{100}}$} & \textbf{$F_{\text{avg}}^{\text{tel}}$} & \textbf{$\langle F_{\text{avg}}^{\text{tel}}\rangle _{T_1,...,T_{100}}$} & \textbf{$F_{\text{avg}}^{\text{tel}}$} & \textbf{$\langle F_{\text{avg}}^{\text{tel}}\rangle _{T_1,...,T_{100}}$} \\
					\hline
					$N = 15$ & 0.607 & 0.606 & 0.662 & 0.66 & 0.56 & 0.577 & 0.577 & 0.576 \\
					\hline
					$N = 127$ & 0.515  & 0.515 & 0.593 & 0.593 & 0.508 & 0.511 & 0.512 & 0.512\\
					\hline
				\end{tabular}
			}
			\caption{Comparison of $F_{\text{avg}}^{\text{tel}}$ and $\langle F_{\text{avg}}^{\text{tel}}\rangle _{T_1,...,T_{100}}$ for $N=15$ and $127$ for different types of Binary trees.}
			\label{tab:uniform2}
		\end{table}
		
		
		\section{Conclusion} \label{sec6}
		
		This article presents a comprehensive analytical investigation into the ability of hierarchical binary networks to transmit quantum information via teleportation. We consider four distinct network topologies - the directed and undirected asymmetric binary trees (DABT, UABT) and the directed and undirected symmetric binary trees (DSBT, USBT) - to understand the influence of directionality and symmetry on network performance. The quality of quantum information transfer is quantified by the average of the maximum teleportation fidelity, $F^{\text{tel}}_{\text{avg}}$.
		By modeling the network edges with simple Werner states, we derive analytical expressions for $F^{\text{tel}}_{\text{avg}}$
		for each network type and explore their limiting values as the number of nodes, $N$, approaches infinity (similar to the fractal tree).  This analytical framework allows us to precisely identify the parameter ranges for which these networks can exhibit a quantum advantage, defined by $F^{\text{tel}}_{\text{avg}}>2/3$. 
		Furthermore, we explore the strategic role of adding maximally entangled states to the network to enhance this quantum advantage. To bridge the gap between idealized models and real-world systems, we also investigate the average teleportation fidelity in a more general setting where link parameters are heterogeneous and chosen from a uniform distribution. There are numerous techniques to reduce the loss and decoherence in the fibres, such as using low-loss fibre, improving coupling and optics, and stabilising the phase and polarization. Alongside this, the quantum error correction techniques, such as the Quantum Parity Code (QPC) \cite{Yan2022, Yan25}, and the Calderbank-Shor-Steane (CSS) code \cite{muralidharan14}, can be used to improve the communication efficiency due to the resilience to photon loss.
		Our work provides a rigorous foundation for identifying resourceful tree networks for quantum information transmission (teleportation). The analytical expressions obtained, along with the detailed exploration of large-$N$ convergence and the roles of symmetry and directionality, contribute to a deeper understanding of quantum network topologies. In this process, we provide a generalized method for the analytical calculation of pathlengths in all four graph types, a methodological advancement that can be applied to more general Cayley tree structures.
		\\
		
		\bmhead{Acknowledgement}
		S.R. would like to acknowledge DST-India for the INSPIRE Fellowship (IF220695) support. The authors also acknowledge  Kishore Kothapalli for having useful discussions, which helped to improve the quality of the manuscript.
		\bibliography{reference}

		\newpage
		
		\begin{appendices}
			
			\section{Average teleportation fidelity of binary tree}
			
			\subsection{Directed  Binary Trees} \label{app:A}
			
			\noindent \textbf{Directed Asymmetric Binary Trees (DABT):}\\
			
			\noindent Let $l_d^{(r)}$ denote the number of $r$-length paths in a DABT with depth $d$. If $r>d$, then there is no path of length $r$ as the tree is directed. So, $l^{(r)}=0$ for $r>d$. For $r=1$, $l_d^{(1)}$ counts the number of edges. Hence, $l_d^{(1)}=2d$. For $r\geq 2$, $l_d^{(r)}$ counts the number of $r$-length paths in the DABT of depth $d-1$ rooted at the vertex $\circled{2}$ and the additional 2 paths that include the vertex $\circled{1}$, namely, $(1,2,\cdots,2r-2,2r)$ and $(1,2,\cdots,2r-2,2r+1)$. Thus, for a fixed $r$ the relation we get is $l_d^{(r)}=l_{d-1}^{(r)}+2$, with the base case $l_{r}^{(r)}=2$, since, in the DABT with depth $r$, the $r$-length paths are $(1,2,\cdots,2r-2,2r)$ and $(1,2,\cdots,2r-2,2r+1)$. Hence, the $d$-th term of this arithmetic progression we get is $l_d^{(r)}=2+2(d-r).$ Therefore, combining the above cases, we get
			\begin{equation}
				l_d^{(r)}=
				\begin{cases}
					2+2(d-r),& \text{if } r\leq d\\
					0,& \text{otherwise}.
				\end{cases}
				\label{eq:A0}
			\end{equation}

			\noindent Hence, for $\mathcal{F}_r = \frac{1+p^r}{2}$, the $F^{\text{tel}}_{\text{avg}}$ in the case of DABT is
			\begin{eqnarray}
				\left.F^{\text{tel}}_{\text{avg}} (d,p)\right|_{\text{DABT}} &=& \frac{\sum\limits_{r=1}^{d} l_d^{(r)}\mathcal {F}_r}{\sum\limits_{r=1}^{d} l_d^{(r)}}=
				\frac{\sum\limits_{r=1}^{d} [2+2(d-r)]\left(\frac{1+p^r}{2}\right)}{\sum\limits_{r=1}^{d} [2+2(d-r)]}=\frac{\sum\limits_{r=1}^{d} [1+(d-r)](1+p^r)}{d(d+1)} \nonumber \\
				&=&\frac{1}{2}+\frac{p[p(p^d-1)+d(1-p)]}{d(d+1)(1-p)^2} 
				= \frac{1}{2}+\varepsilon(d).
				\label{DABTpluserror}
			\end{eqnarray}
			Let the number of nodes of a DABT of depth $d$ be $N=N(d)$, then for this case, $N(d) = 2d+1$. Hence, we can write $F^{\text{tel}}_{\text{avg}}$ in terms of $N$ as
			\begin{eqnarray}
				\left.F^{\text{tel}}_{\text{avg}} (N,p)\right|_{\text{DABT}} 
				= \frac{1}{2}+\frac{2p[2p(p^{\frac{N-1}{2}}-1)+(N-1)(1-p)]}{(N-1)(N+1)(1-p)^2}.
			\end{eqnarray}
			\noindent It is easy to see that $\varepsilon=\mathcal{O}\left(\frac{1}{d}\right)=\mathcal{O}\left(\frac{1}{N}\right)$ for this case.\\
			\begin{figure}[h]
				\centering
				\includegraphics[scale = 0.20]{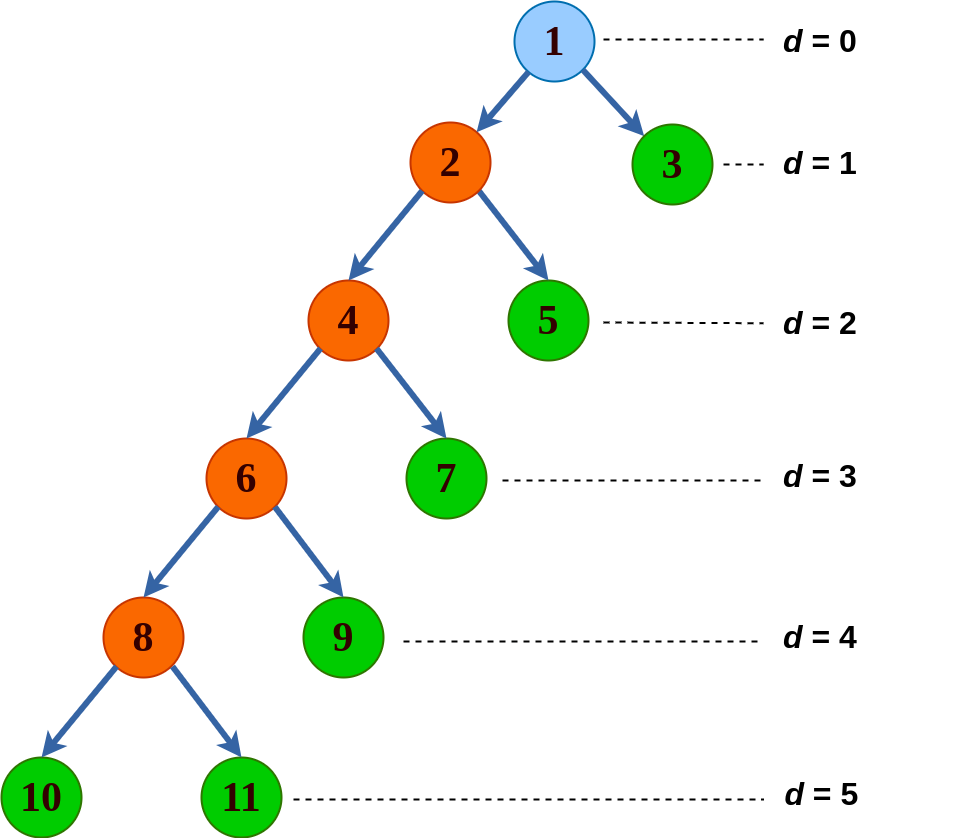}
				\caption{A directed asymmetric binary tree with 11 Nodes. The depth of the tree is 5.}
				\label{DABT_h5}
			\end{figure} 
			%
			\par \noindent \textbf{Directed Symmetric Binary Trees (DSBT)}:\\
			\noindent We denote the number of $r$-length paths starting from the base vertex $\circled{1}$ by $h_r$. At each level, there are two choices for selecting a vertex to include in the path up to depth $r$. Thus, $h_r = 2^r$.
			Now, let us denote by $l_d^{(r)}$ the number of $r$-length paths in a DSBT with depth $d$. If $r > d$, then there is no $r$-length path as this is directed tree, so $l_d^{(r)} = 0$  for $r>d$.\\
			
			\noindent For $r \leq d$, $l_d^{(r)}$ counts the total number of paths of length $r$ in the subtrees rooted at the vertices $\circled{2}$ and $\circled{3}$ along with the paths that start with the base vertex $\circled{1}$. So, for a fixed $r$ we can write the recurrence for $\{l_d^{(r)}\}_{d}$ in this form
			
			\begin{equation}\label{A2}
				l_d^{(r)} = 2l_{d-1}^{(r)} + h_r=2l_{d-1}^{(r)} + 2^r.  
			\end{equation} 
			Now, let $z_d^{(r)} = l_d^{(r)} + 2^r$. Then Eq. (\ref{A2}) becomes
			$z_d^{(r)}= 2 z_{d - 1}^{(r)}$,
			which has the general solution
			$z_d^{(r)} = 2^d s^{(r)} $ \cite{RosenDiscreteMath}
			where, $s^{(r)}$ is independent of $d$. Therefore, we get
			\begin{equation}\label{A3}
				l_d^{(r)} = 2^d s^{(r)} - 2^r.
			\end{equation}
			If we take $d=r$, then we will have $l_r^{(r)} = 2^r$, as it will only count all the paths starting with the base vertex $\circled{1}$. Thus, using this in Eq. (\ref{A3}), we obtain $s^{(r)} = 2$ for all $r \leq d$. Therefore, for $r \leq d$, we will have
			\begin{equation}\label{A4}
				l_d^{(r)} = 2^{d+1} - 2^r.
			\end{equation}
			\noindent Therefore, combining the above cases, we get
			\begin{equation}
				l_d^{(r)}=
				\begin{cases} 
					2^{d+1} - 2^r,& \text{if} ~r\leq d\\
					0,& \text{otherwise}.
				\end{cases}
				\label{eq:A4}
			\end{equation}

			\noindent Hence, for $\mathcal{F}_r = \frac{1+p^r}{2}$, the $F^{\text{tel}}_{\text{avg}}$ in the case of DSBT is
			\begin{align}
				\left.F^{\text{tel}}_{\text{avg}} (d,p)\right|_{\text{DSBT}} &= \frac{\sum\limits_{r=1}^{d} l_d^{(r)}\mathcal {F}_r}{\sum\limits_{r=1}^{d} l_d^{(r)}} = 
				\frac{\sum\limits_{r=1}^{d} [2^{d+1} - 2^r]\left(\frac{1+p^r}{2}\right)}{\sum\limits_{r=1}^{d} [2^{d+1} - 2^r]}
				=\frac{1}{2}+\frac{1}{2}\cdot \frac{\sum\limits_{r=1}^{d} [2^{d+1} - 2^r]\cdot p^r}{\sum\limits_{r=1}^{d} [2^{d+1} - 2^r]} \nonumber \\
				&=
				\begin{cases}
					\displaystyle
					\frac{1}{2} +\frac{2^{d+1} - d - 2}{2\left[(d - 1) \cdot 2^{d+1} + 2\right]} , & \text{if }~ p = \frac{1}{2}, \\[12pt]
					\displaystyle
					\frac{1}{2} +\frac{p\left[(2^d - 1) - p(2^{d+1}-1) + 2^d p^{d+1}\right]}{2(1-p)(1-2p)\left[(d - 1)\cdot 2^d + 1\right]}, & \text{otherwise},
				\end{cases}\\
				&= \frac{1}{2}+\varepsilon(d).
				\label{DSBTpluserror}
			\end{align}
			If the total number of nodes of a DSBT of depth $d$ is $N=N(d)$, then $N(d) = 2^{d+1}-1$. Hence, we can write $F^{\text{tel}}_{\text{avg}}$ in terms of $N$ as
			\begin{eqnarray}
				\left.F^{\text{tel}}_{\text{avg}} (N,p)\right|_{\text{DSBT}} 
				&=& \begin{cases}
					\displaystyle
					\frac{1}{2} +\frac{N-\log_2(N+1)}{2\left[(N+1)\log_2(N+1)-2N \right]} , & \text{if }~ p = \frac{1}{2}, \\[12pt]
					\displaystyle
					\frac{1}{2}+\frac{p[N(1-2p)+(N+1)p^{\log_2(N+1)}-1]}{4(1-p)(1-2p)\left[(N+1)\log_2(N+1)-2N \right]}, & \text{otherwise}.
				\end{cases}
			\end{eqnarray}
			So, in this case, $\varepsilon=\mathcal{O}\left(\frac{1}{d}\right)=\mathcal{O}\left(\frac{1}{\log_2 N}\right)$ for this case.\\
			
			\begin{figure*}[h!]
				\centering
				\includegraphics[scale = 0.18]{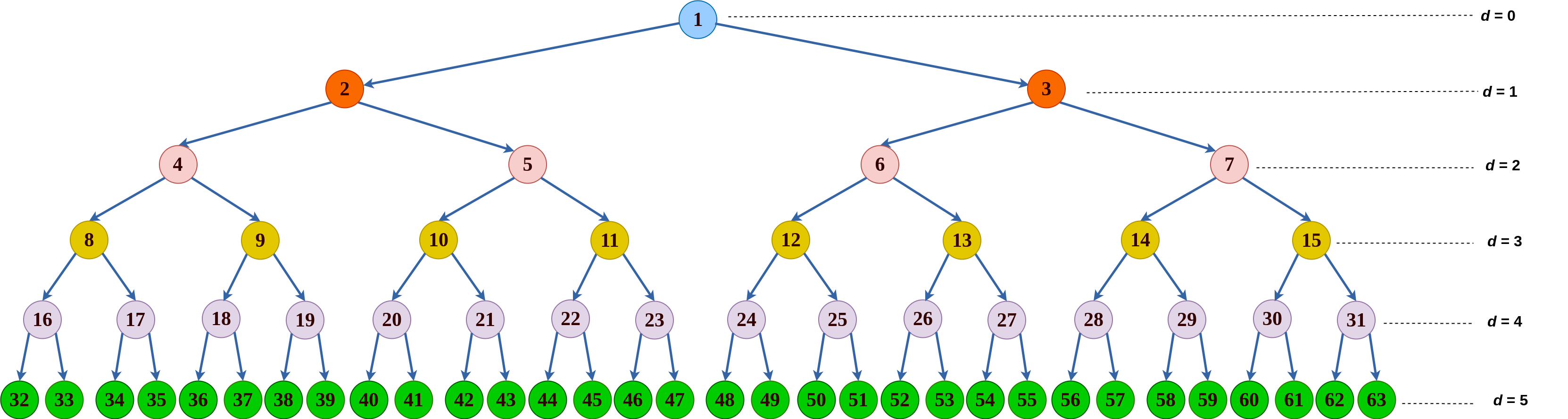}
				\caption{A directed symmetric binary tree with 63 Nodes. The depth of the tree is 5.}
				\label{DSBT_63}
			\end{figure*}
			
			
			\subsection{Undirected  Binary Trees}\label{app:B}
			
			\noindent \textbf{Undirected Asymmetric Binary Trees (UABT)}: 
			
			\noindent Let $l_d^{(r)}$ denote the number of $r$-length paths in a UABT with depth $d$. As it is an undirected and asymmetric tree, for $r>d+1$, there is no path of length $r$. So, $l_d^{(r)}=0$ for $r>d+1$. 
			
			\noindent For $r=1$, $l_d^{(1)}$ counts the number of edges. Hence, $l_d^{(1)}=2d$.
			
			\noindent For $r=2$, $l_d^{(2)}$ is the number of $2$-length paths in the UABT of depth $d-1$ rooted at the vertex $\circled{2}$ and the additional 3 paths that includes the vertex $\circled{1}$, namely, $(3,1,2)$, $(1,2,4)$ and $(1,2,5)$. Hence, we get the relation $l_d^{(2)}=l_{d-1}^{(2)}+3$, with the base case $l_1^{(2)}=1$. Therefore, the $d$-th term of this arithmetic progression is $l_d^{(2)}=1+3(d-1).$ \\
			\noindent Similarly, for $r\geq 3$, $l_d^{(r)}$ counts the number of $r$-length paths in the UABT of depth $d-1$ rooted at the vertex $\circled{2}$ and the additional 4 paths that includes the vertex $\circled{1}$, namely, $(3,1,2,\cdots,2r-2)$, $(3,1,2,\cdots,2r-1)$, $(1,2,4,\cdots,2r)$ and $(1,2,4,\cdots,2r+1)$. Thus, the relation we get is $l_d^{(r)}=l_{d-1}^{(r)}+4$, with the base case $l_{r-1}^{(r)}=2$, since, in the UABT with depth $r-1$, the $r$-length paths are $(3,1,2,\cdots,2r-2)$ and $(3,1,2,\cdots,2r-1)$. Therefore, the $d$-th term of this arithmetic progression we get is $l_d^{(r)}=2+4(d-r+1).$ Therefore, combining the above cases, we get
			\begin{eqnarray}
				l_d^{(r)}=
				\begin{cases}
					2d,& \text{if} ~r=1\\
					1+3(d-1),& \text{if} ~r=2\\
					2+4(d-r+1),& \text{if}  ~3\leq r\leq d+1\\
					0,& \text{otherwise}.
				\end{cases}
				\label{eq:A5}
			\end{eqnarray}

			\noindent Hence, for $\mathcal{F}_r = \frac{1+p^r}{2}$, the $F^{\text{tel}}_{\text{avg}}$ in the case of UABT is $\frac{3+2p+p^2}{6}$ for $d=1$ and for $d\geq 2$,
			\begin{align}
				\left.F^{\text{tel}}_{\text{avg}}(d,p)\right|_{\text{UABT}} &=  \frac{\sum\limits_{r=1}^{d} l_d^{(r)}\mathcal {F}_r}{\sum\limits_{r=1}^{d} l_d^{(r)}} 
				=\frac{1}{2}+\frac{1}{2}\cdot \frac{\left[2dp+\left(1+3(d-1)\right)p^2+\sum\limits_{r=3}^{d+1} \left(2+4(d-r+1)\right)p^r\right]}{\left[2d+\left(1+3(d-1)\right)+\sum\limits_{r=3}^{d+1} \left(2+4(d-r+1)\right)\right]} \nonumber\\
				&=\frac{1}{2}+\frac{\sum\limits_{r=1}^{d+1} \left(6+4(d-r)\right)p^r-(dp^2+2dp+2p)}{2d(2d+1)} \nonumber\\
				&=\frac{1}{2}+\frac{2dp - (d + 2)p^2 - 2p^3 - dp^4 + 2p^{d+2} + 2p^{d+3}}{2d(2d + 1)(1 - p)^2}\\
				&= \frac{1}{2}+\varepsilon(d).
				\label{UABTpluserror}
			\end{align}
			
			\noindent The number of nodes of a UABT of depth $d$ be $N=N(d)$, then $N(d) = 2d+1$. Hence, the $F^{\text{tel}}_{\text{avg}}$ in case of UABT is
			\begin{eqnarray}
				&&\left.F^{\text{tel}}_{\text{avg}} (N,p)\right|_{\text{UABT}} \nonumber\\
				&=& \frac{1}{2}+\frac{2p(N-1)-(N+3)p^2-4p^3-(N-1)p^4+4p^{\frac{N+3}{2}}+4p^{\frac{N+5}{2}}}{2N(N-1)(1-p)^2},
			\end{eqnarray}
			where $\mathcal F_i \equiv (1+p^i)/2$. Thus, $\varepsilon=\mathcal{O}\left(\frac{1}{d}\right)=\mathcal{O}\left(\frac{1}{N}\right)$ for this case.
			
			\begin{figure}[h!]
				\centering
				\includegraphics[scale = 0.20]{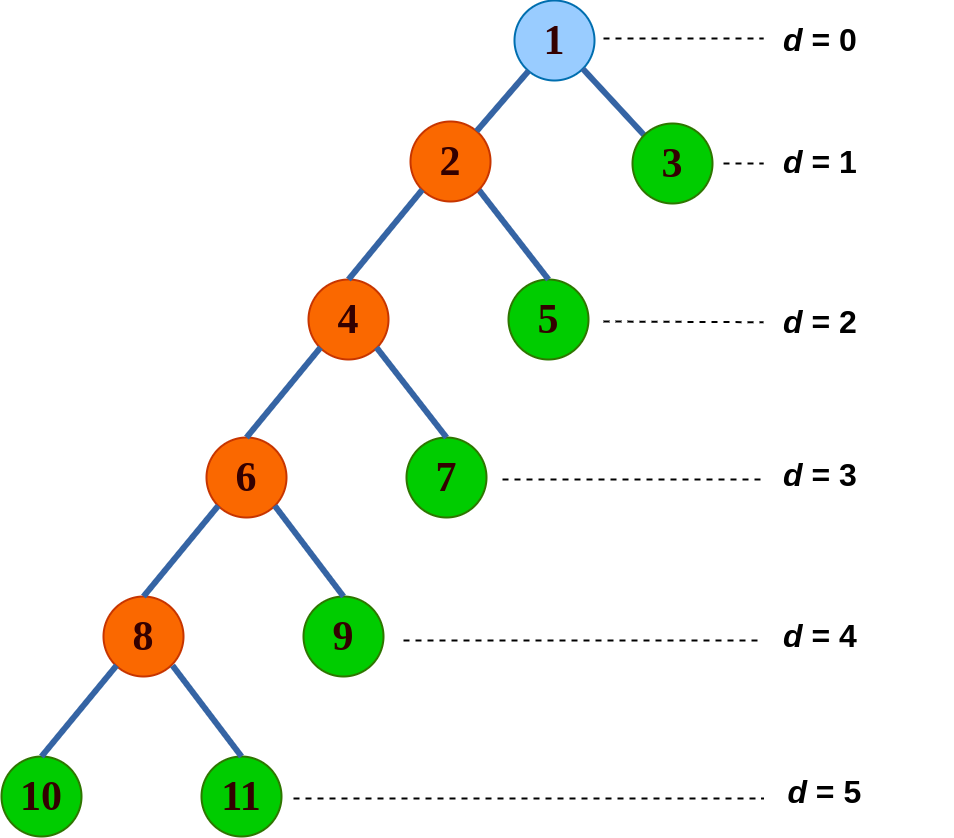}
				\caption{A undirected asymmetric binary tree with 11 Nodes. The depth of the tree is 5.}
				\label{UABT_11}
			\end{figure}
			
			\newpage
			\noindent \textbf{Undirected  Symmetric Binary Trees (USBT)}:\\
			\noindent Here again, $h_r$ denotes the number of $r$-length paths starting from the base vertex $\circled{1}$, and at each level, there are two choices for selecting a vertex to include in the path up to depth $r$. So, $h_r = 2^r$. Now, let us denote by $l_d^{(r)}$ the number of $r$-length paths in a USBT with depth $d$. In this case, for $r > 2d$, there is no $r$-length path as it is an undirected and symmetric tree, and hence $l_d^{(r)} = 0$  for $r>2d$.
			
			\noindent Now, for $1\leq r \leq 2d$, $l_d^{(r)}$ counts the total number of paths of length $r$ in the subtrees rooted at the vertices $\circled{2}$ and $\circled{3}$ along with the paths that include the base vertex $\circled{1}$. So, for a fixed $r$ we can write the recurrence for $\{l_d^{(r)}\}_{d}$ in this form
			\begin{equation}
				l_d^{(r)} = 2l_{d-1}^{(r)} + c_d^{(r)},  
			\end{equation}
			where, $c_d^{(r)}$ denotes the number of paths that include the base vertex $\circled{1}$, while connecting the two subtrees rooted at vertices $\circled{2}$ and $\circled{3}$. We consider the following cases:\\
			
			\noindent \textbf{Case I:} $d < r \leq 2d$:\\
			In this case, no path begins with the base vertex $\circled{1}$. Hence, $c_d^{(r)}$ counts the number of paths where vertex $\circled{1}$ is not an endpoint. That is, it counts the total number of paths with $k$ edges on the left and $r-k$ edges on the right of vertex $\circled{1}$, for all $r-d \leq k \leq d$. Therefore,
			$$ c_d^{(r)}= \sum_{k = r - d}^{d} h_{k - 1}\cdot h_{r - k - 1}
			= \sum_{k = r - d}^{d} 2^{k-1}\cdot 2^{r-k-1}
			= (2d - r + 1)\cdot 2^{r - 2}.$$  
			Thus, we obtain the recurrence to be
			\begin{equation}\label{2}
				l_d^{(r)} = 2l_{d-1}^{(r)} + (2d - r + 1)\cdot 2^{r - 2}.
			\end{equation}
			
			\noindent Let $a_d^{(r)} = 2^{-(r - 2)} l_d^{(r)} + (2d - r + 5)$. Then Eq.\ (\ref{2}) becomes
			$a_d^{(r)}= 2 a_{d - 1}^{(r)}$ ,
			which has the general solution
			$a_d^{(r)} = 2^d \cdot p^{(r)} $ \cite{RosenDiscreteMath}
			where, $p^{(r)}$ is independent of $d$. Therefore, we get
			\begin{equation}\label{3}
				l_d^{(r)} = \left[2^d p^{(r)} - (2d - r + 5)\right]\cdot 2^{r - 2}.
			\end{equation}
			If $r = 2k$, then $l_{k-1}^{(2k)} = 0$, so $l_k^{(2k)} = 2^{2k - 2}$. Using this in Eq.\ (\ref{3}), we get $p^{(2k)} = 3 \cdot 2^{-(k - 1)}$. Therefore,
			\begin{equation}\label{4}
				{l_d^{(2k)} = 3 \cdot 2^{d + k - 1} - (2d - 2k + 5) \cdot 2^{2k - 2}}.    
			\end{equation}
			Now, if $r = 2k - 1$, then $
			l_{k - 1}^{(2k - 1)} = 0$, so $l_k^{(2k - 1)} = 2^{2k - 2}$. Hence, again from Eq.\ (\ref{3}), we get $p^{(2k - 1)} = 2^{-(k - 3)}$. \noindent Therefore,
			\begin{equation}\label{5}
				l_d^{(2k - 1)} = 2^{d + k} - (d - k + 3) \cdot 2^{2k - 2}.
			\end{equation}
			\textbf{Case II:}  $r\leq d$\\
			\noindent In this case, $c_d^{(r)}$ counts the number of paths where vertex $\circled{1}$ is indeed an endpoint and also the number of paths with $k$ edges on the left and $r-k$ edges on the right of vertex $\circled{1}$, for all $1 \leq k \leq r-1$. Thus,
			$$ c_d^{(r)}= h_r+\sum_{k = 1}^{r-1} h_{k - 1} \cdot h_{r - k - 1}
			= 2^r+\sum_{k = 1}^{r-1} 2^{k - 1} \cdot 2^{r - k - 1}
			=(r + 3)\cdot 2^{r - 2}.$$
			
			\noindent  Therefore, the recurrence that we obtain this time is
			\begin{equation}\label{6}
				l_d^{(r)} = 2l_{d-1}^{(r)} + (r + 3) \cdot 2^{r - 2}.
			\end{equation}
			Now, let $b_d^{(r)} = l_d^{(r)} + (r + 3) \cdot 2^{r - 2}$. Then Eq.\ (\ref{6})
			$b_d^{(r)}= 2 b_{d - 1}^{(r)}$,
			which has the general solution
			$b_d^{(r)} = 2^d q^{(r)} $ \cite{RosenDiscreteMath}
			where, $q^{(r)}$ is independent of $d$. Therefore, we get
			\begin{equation}\label{7}
				l_d^{(r)} = 2^d q^{(r)} - (r + 3)\cdot 2^{r-2}.
			\end{equation}
			If $r = 2k$, then by taking $d=2k-1$ in Eq.\ (\ref{4}), we get 
			$$l_{2k - 1}^{(2k)} = 3 \cdot 2^{3k - 2} - (2k + 3) \cdot 2^{2k - 2}.$$
			
			\noindent  Thus, using this in Eq.\ (\ref{7}), we obtain $q^{(2k)} = 3 \cdot 2^{k - 1}$. Therefore,
			\begin{equation}\label{8}
				l_d^{(2k)} = 3 \cdot 2^{d + k - 1} - (2k + 3) \cdot 2^{2k - 2}.
			\end{equation}
			Now, if $r = 2k + 1$, then again from Eq.\ (\ref{5}), we get 
			$$l_{2k}^{(2k + 1)} = 2^{3k + 1} - (k + 2) \cdot 2^{2k}.$$
			
			\noindent Thus, using this in Eq.\ (\ref{7}), we obtain $q^{(2k + 1)} = 2^{k + 1}$. Therefore,
			\begin{equation}\label{9}
				l_d^{(2k + 1)} = 2^{d + k + 1} - (k + 2) \cdot 2^{2k}.
			\end{equation}
			\noindent Therefore, combining all four cases, we get
			\begin{equation}
				l_d^{(r)}=
				\begin{cases}
					3 \cdot 2^{d + (\frac{r}{2}) - 1} - (2d-r+5) \cdot 2^{r - 2},& \text{if} ~d<r\leq 2d ~ \text{and} ~r~ \text{is even}\\
					2^{d + (\frac{r+1}{2})} - (2d-r+5) \cdot 2^{r - 2},& \text{if} ~d<r\leq 2d ~\text{and} ~r ~\text{is odd}\\
					3 \cdot 2^{d + (\frac{r}{2}) - 1} - (r+3) \cdot 2^{r - 2},& \text{if} ~r\leq d ~\text{and} ~r ~\text{is even}\\
					2^{d + (\frac{r+1}{2})} - (r+3) \cdot 2^{r - 2}, & \text{if} ~r\leq d ~ \text{and} ~ r ~ \text{is odd}\\
					0,& \text{otherwise}.
				\end{cases}
			\end{equation}
			
			\noindent Hence, for $\mathcal{F}_r = \frac{1+p^r}{2}$, the $F^{\text{tel}}_{\text{avg}}$ in the case of USBT is\\
			\resizebox{\textwidth}{!}{$
				\begin{aligned}
					&\left.F^{\text{tel}}_{\text{avg}} (d,p)\right|_{\text{USBT}} \\
					&=
					\frac{\left[
						\begin{aligned}
							&\sum\limits_{\substack{r \leq d \\ r~\text{even}}}
							\left(3 \cdot 2^{d + \frac{r}{2} - 1} - (r+3) \cdot 2^{r - 2}\right) \cdot \left(\frac{1+p^r}{2}\right) 
							+ \sum\limits_{\substack{r \leq d \\ r~\text{odd}}}
							\left(2^{d + \frac{r+1}{2}} - (r+3) \cdot 2^{r - 2}\right) \cdot \left(\frac{1+p^r}{2}\right)+ \\
							& \sum\limits_{\substack{d < r \leq 2d \\ r~\text{even}}}
							\left(3 \cdot 2^{d + \frac{r}{2} - 1} - (2d - r + 5) \cdot 2^{r - 2}\right) \cdot \left(\frac{1+p^r}{2}\right) 
							+ \sum\limits_{\substack{d < r \leq 2d \\ r~\text{odd}}}
							\left(2^{d + \frac{r+1}{2}} - (2d - r + 5) \cdot 2^{r - 2}\right) \cdot \left(\frac{1+p^r}{2}\right)
						\end{aligned}\right]
					}
					{\left[
						\begin{aligned}
							&\sum\limits_{\substack{r \leq d \\ r~\text{even}}}
							\left(3 \cdot 2^{d + \frac{r}{2} - 1} - (r+3) \cdot 2^{r - 2}\right) 
							+ \sum\limits_{\substack{r \leq d \\ r~\text{odd}}}
							\left(2^{d + \frac{r+1}{2}} - (r+3) \cdot 2^{r - 2}\right)+ \\
							& \sum\limits_{\substack{d < r \leq 2d \\ r~\text{even}}}
							\left(3 \cdot 2^{d + \frac{r}{2} - 1} - (2d - r + 5) \cdot 2^{r - 2}\right) 
							+ \sum\limits_{\substack{d < r \leq 2d \\ r~\text{odd}}}
							\left(2^{d + \frac{r+1}{2}} - (2d - r + 5) \cdot 2^{r - 2}\right)
						\end{aligned}\right]
					} \\
					&=\frac{1}{2}+\frac{1}{2}
					\frac{
						\sum\limits_{\substack{r =1 \\ r~\text{even}}}^{2d}
						3 \cdot 2^{d + \frac{r}{2} - 1} \cdot p^r+
						\sum\limits_{\substack{r =1 \\ r~\text{odd}}}^{2d}
						2^{d + \frac{r+1}{2}} \cdot p^r-
						\sum\limits_{r=1}^{d} (r+3) \cdot 2^{r - 2}\cdot p^r-
						\sum\limits_{r=d+1}^{2d}(2d - r + 5) \cdot 2^{r - 2}\cdot p^r
					}{
						\sum\limits_{\substack{r =1 \\ r~\text{even}}}^{2d}
						3 \cdot 2^{d + \frac{r}{2} - 1} +
						\sum\limits_{\substack{r=1 \\ r~\text{odd}}}^{2d}
						2^{d + \frac{r+1}{2}}-
						\sum\limits_{r=1}^{d} (r+3) \cdot 2^{r - 2}-
						\sum\limits_{r=d+1}^{2d}(2d - r + 5) \cdot 2^{r - 2}
					} \\
					&=
					\frac{1}{2}+\frac{1}{2}
					\frac{
						\sum\limits_{k=1}^{d}
						3 \cdot 2^{d + k - 1} \cdot p^{2k}+
						\sum\limits_{k=1}^{d}2^{d + k} \cdot p^{2k-1}
						-\sum\limits_{r=1}^{d} (r+3) \cdot 2^{r - 2}\cdot p^r
						-\sum\limits_{t=1}^{d}(d-t + 5) \cdot 2^{d+t - 2}\cdot p^{d+t}
					}{
						\sum\limits_{k=1}^{d}
						3 \cdot 2^{d + k - 1}+
						\sum\limits_{k=1}^{d}2^{d + k}
						-\sum\limits_{r=1}^{d} (r+3) \cdot 2^{r - 2}
						-\sum\limits_{t=1}^{d}(d-t + 5) \cdot 2^{d+t - 2}
					} \\
					&=
					\frac{1}{2}+\frac{1}{2}
					\frac{
						\left(3\cdot 2^{d-1}+\frac{2^d}{p}\right)\sum\limits_{k=1}^{d}(2p^2)^k
						-\left(\frac{3}{4}+(d+5)\cdot 2^{d-2}\cdot p^d\right) \sum\limits_{k=1}^{d}(2p)^k
						+ \left( 2^{d-2}\cdot p^d-\frac{1}{4}\right) \sum\limits_{k=1}^{d}k(2p)^k
					}{
						\left(3\cdot 2^{d-1}+2^d\right)\sum\limits_{k=1}^{d}2^k
						-\left(\frac{3}{4}+(d+5)\cdot 2^{d-2}\right) \sum\limits_{k=1}^{d}2^k
						+ \left( 2^{d-2}-\frac{1}{4}\right) \sum\limits_{k=1}^{d}k2^k
					} \\
					&=
					\begin{cases}
						\displaystyle
						\frac{1}{2} + \frac{14(2^d - 1) - d(d + 8)}{8(2^d - 1)(2^{d+1} - 1)}, & \text{if }~ p = \frac{1}{2},\\[12pt]
						\displaystyle
						\frac{1}{2} + \frac{d \cdot 2^{d+1} - \sqrt{2}(2^{d/2} - 1)\left[(\sqrt{2} - 1)\left\{3 + 5 \cdot 2^{d/2}\right\} + (2^{d/2} - 1)\right]}{8(3 - 2\sqrt{2})(2^d - 1)(2^{d+1} - 1)}, & \text{if }~ p = \frac{1}{\sqrt{2}}, \\[12pt]
						\displaystyle
						\frac{1}{2}
						+ \frac{2^{d-1}p(3p + 2)\left((2p^2)^d - 1\right)}{(2p^2 - 1)(2^d - 1)(2^{d+1} - 1)}
						- \frac{p\left((2p)^d - 1\right)\left[(2p - 1)\left\{3 + 5 \cdot (2p)^d\right\} + \left((2p)^d - 1\right)\right]}{4(2p - 1)^2(2^d - 1)(2^{d+1} - 1)}, & \text{otherwise}.
					\end{cases}
				\end{aligned}
				$}
			
			\par \noindent Therefore, we can write,
			\begin{eqnarray}
				\left.F^{\text{tel}}_{\text{avg}} (d,p)\right|_{\text{USBT}} = \frac{1}{2}+\varepsilon(d). \label{USBTpluserror}
			\end{eqnarray}
			%
			%
			\noindent Here, the total number of nodes of a USBT of depth $d$ is $N=N(d)$, then $N(d) = 2^{d+1}-1$. Hence, the $F^{\text{tel}}_{\text{avg}}$ in case of USBT in terms of $N$ can be written by substituting each $d$ with $\log_2[(N+1)/2]$ and thus for this case,
			$$\varepsilon=\begin{cases}
				\mathcal{O}\left(2^{-d}\right)=\mathcal{O}\left(\dfrac{1}{N}\right), & \text{if } 0 < p < \dfrac{1}{\sqrt{2}}, \\[12pt]
				\mathcal{O}\left(d \cdot 2^{-d}\right)=\mathcal{O}\left(\dfrac{\log_2 N}{N}\right), & \text{if } p = \dfrac{1}{\sqrt{2}}, \\[12pt]
				\mathcal{O}\left(p^{2d}\right)=\mathcal{O}\left(N^{-2\log_2 (1/p)}\right), & \text{if } \dfrac{1}{\sqrt{2}} < p < 1.
			\end{cases}$$
			
			We did not consider the case $p=1$ while computing the average fidelity in the above four trees, namely DABT, DSBT, UABT and USBT, since then we would have $\mathcal{F}_r = \frac{1+p^r}{2}=1$, hence
			$$F^{\text{tel}}_{\text{avg}}=\frac{\sum\limits_{r} l_d^{(r)}\mathcal {F}_r}{\sum\limits_{r}l_d^{(r)}} 
			=1.$$
			\begin{figure}[h!]
				\centering
				\includegraphics[scale = 0.18]{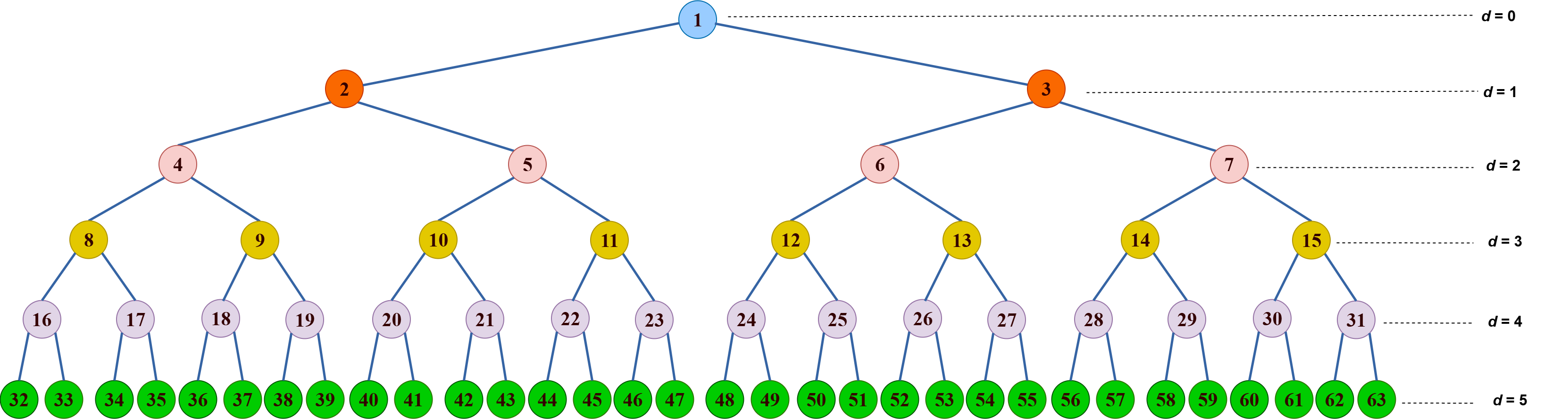}
				\caption{A undirected symmetric binary tree with 63 Nodes. The depth of the tree is 5.}
				\label{USBT_63}
			\end{figure}
			%
			%
			%
			%
			%
			%
			%
			
			
			
			
			\newpage
			\section{Large $N$ limit}\label{app2}
			To understand the large $N$ behaviour for all the cases of binary trees, we consider the formula of $F_{\mathrm{avg}}^{\mathrm{tel}}$. For any value of $p<1$, we can write the formula of $F_{\mathrm{avg}}^{\mathrm{tel}}(d,p)$ in terms of $\frac{1}{2}+\varepsilon(d)$, where $\varepsilon(d)$ is the additional term (See Eq. \eqref{DABTpluserror}, \eqref{DSBTpluserror}, \eqref{UABTpluserror}, and \eqref{USBTpluserror}).
			This term converges to 0 when $d$ is large. To observe the behaviour, we have plotted $\varepsilon(N)$ (which can be obtained by simply changing the $d$ with $N$ for each symmetric and asymmetric trees) vs $N(d)$.
			\begin{figure}[h]
				\centering
				\begin{minipage}{0.23\linewidth}
					\centering
					\includegraphics[scale = 0.19]{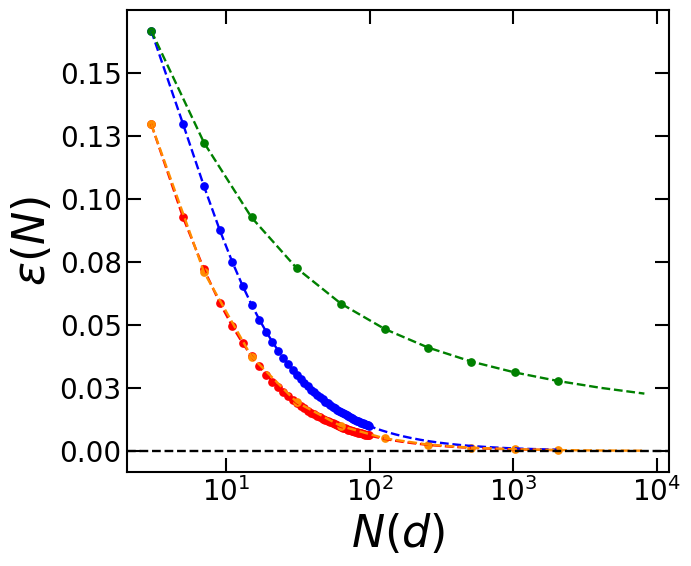}\\
					\qquad(a)
				\end{minipage}
				\hfill
				\begin{minipage}{0.23\linewidth}
					\centering
					\includegraphics[scale = 0.19]{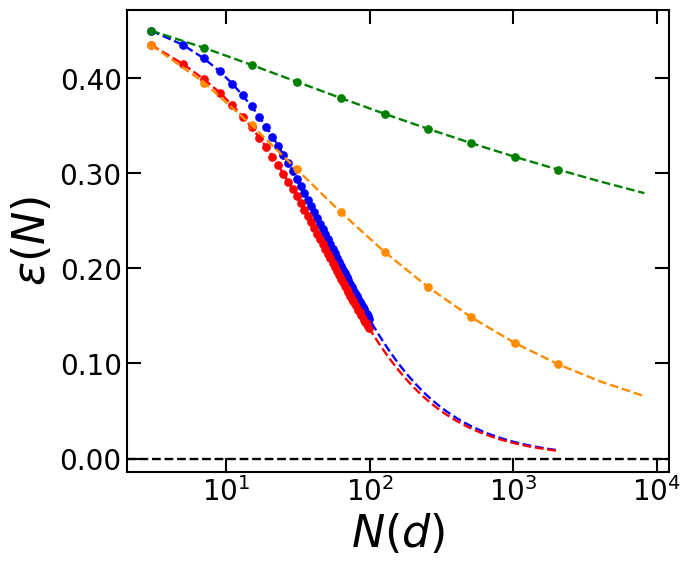}\\
					\qquad(b)
				\end{minipage}
				\hfill
				\begin{minipage}{0.23\linewidth}
					\centering
					\includegraphics[scale = 0.19]{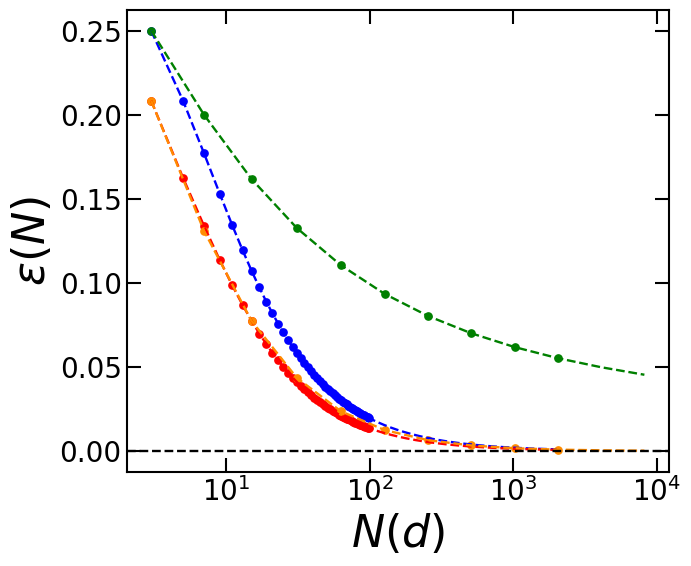}\\
					\qquad(c)
				\end{minipage}
				\hfill
				\begin{minipage}{0.23\linewidth}
					\centering
					\includegraphics[scale = 0.19]{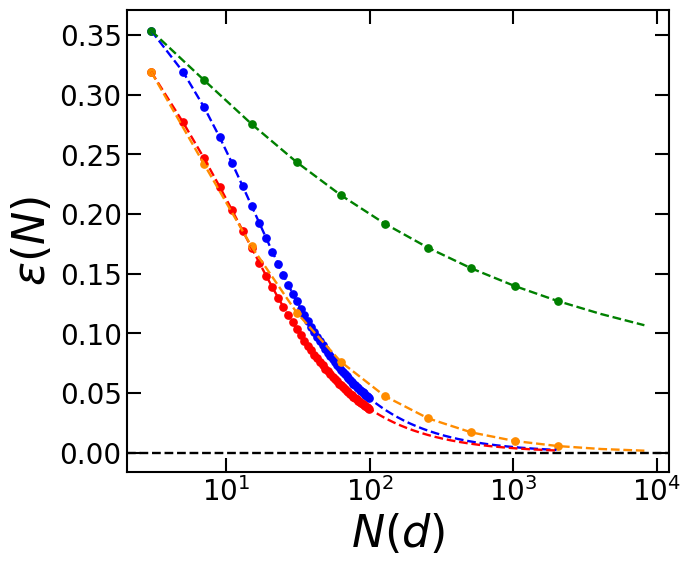}\\
					\qquad(d)
				\end{minipage}
				\caption{ The plot of $\varepsilon(N)$ vs $N(d)$ in the case of all four types of binary trees (a) for $p=0.333$, (b) for $p=0.9$, (c) for $p=1/2$, and (d) for $p=1/\sqrt{2}$. Here, the blue, green, red and orange lines denote DABT, DSBT, UABT and USBT, respectively. For all the cases, $\varepsilon(N)$ converges to $0$ when $N(d) $ is large.}
				\label{large N error}
			\end{figure}
			
		\end{appendices}

	\end{document}